\documentclass[a4paper, 12pt]{article}
\usepackage[utf8]{inputenc}

\usepackage{lmodern}

\usepackage{amsmath,setspace,geometry}
\usepackage{amsthm}
\usepackage{amsfonts}
\usepackage[shortlabels]{enumitem}
\usepackage{rotating}
\usepackage{pdflscape}
\usepackage{physics}
\usepackage{graphicx}
\usepackage{bbm}
\usepackage{booktabs}
\usepackage{subfigure}
\usepackage{tabularray}
\usepackage{subcaption}
\usepackage[dvipsnames]{xcolor}
\usepackage{tabularray}
\usepackage{float}
\usepackage{natbib}
\usepackage{graphicx}
\usepackage{rotating}
\usepackage{codehigh}

\usepackage{tabularx,ragged2e,array,calc,amsmath}
\newcolumntype{C}[1]{>{\Centering}p{#1}}
\newcolumntype{L}{>{\raggedright}X}

\usepackage[normalem]{ulem}
\UseTblrLibrary{booktabs}
\UseTblrLibrary{siunitx}

\NewTableCommand{\tinytableDefineColor}[3]{\definecolor{#1}{#2}{#3}}

\usepackage{hyperref}

\usepackage[]{natbib} 
\bibpunct[:]{(}{)}{,}{a}{}{,}
\usepackage[english]{babel}
\usepackage{float}
\usepackage{caption}
\usepackage{subcaption}
\usepackage{booktabs}
\usepackage{pdfpages}
\usepackage{threeparttable}
\usepackage{lscape}
\usepackage{bm}
\usepackage{bbm}

\usepackage{setspace}
\title{{Welfare Effects of Self-Preferencing by a Platform: Empirical Evidence from Airbnb}\thanks{I thank Masanori Tsuruoka,  Chishio Furukawa, and Kyogo Kanazawa for their helpful comments and advice.}}
\author{Kaede Hanazawa\thanks{Yokohama National University, Department of Economics, E-mail: \texttt{kaedehana2@gmail.com}}}

\date{\today}

\begin{document}
\maketitle
\begin{abstract}
    This paper studies the welfare effects of \textit{self-preferencing} by Airbnb, a practice where Airbnb utilizes its pricing algorithm to prioritize maximizing platform-wide commission revenue rather than optimizing individual host revenues.
    To examine this welfare implication, I construct a Bertrand competition model with differentiated products between Airbnb hosts and hotels.
    Using unique data from Tokyo's 23 wards, I estimate the model and conduct counterfactual simulations to evaluate the welfare effects of self-preferencing.
    Counterfactual simulations reveal that self-preferencing reduces social welfare by 5.08\% on average, equivalent to an annual loss of about 14.90\% of Tokyo's vacation rental market size in 2023 while increasing Airbnb's commission revenue by an average of 37.73\%.
    These findings highlight the significant trade-offs between platform-driven revenue optimization and market efficiency, emphasizing the urgent need for competition policy reforms and greater transparency and accountability in platform practices.
\end{abstract}




\section{Introduction}

With the rapid advancement of digital technology, platform industries have emerged as a dominant force in the global economy. 
Remarkably, five of the ten largest firms by market capitalization are platforms.\footnote{The five platforms are Apple, Amazon, Alphabet (Google), Microsoft, and Meta. Source: Nikkei \url{https://www.nikkei.com/marketdata/ranking-us/market-cap-high/}, accessed October 8, 2024.}
In Japan, the consumer e-commerce market was valued at 24.8 trillion yen in 2023, accounting for approximately 4.19\% of the nation's nominal GDP.\footnote{Based on the Ministry of Economy, Trade, and Industry's ``2023 Survey Report on the Electronic Commerce Market” \url{https://www.meti.go.jp/policy/it_policy/statistics/outlook/R5tyousahoukokusho.pdf}, accessed October 8, 2024.} 

In this context, concerns have arisen about the platform \textit{self-preferencing}, where they prioritize their goods and services above others \citep{Cremer2019, kittaka2023self}. A notable example is the European Commission ruling against Google for abusing its dominant market position by manipulating search results to favor its content, resulting in a 2.42 billion euro fine \citep{european_commission_2017}. In response to such practices, competition authorities are considering proactive approaches to regulating self-preferencing and enforcing antitrust laws \citep{DigitalCompetition2019, EuropeanCommission2020a, EuropeanCommission2020b, USHouse2020, japan_fair_trade_commision2021}. 
Given the growing concerns regarding the impacts of self-preferencing, it is crucial to investigate its implications. 
This study therefore addresses the following research question: “What is the impact of platform self-preferencing on social welfare?”

To examine the research question, I specifically focus on Airbnb and its self-preferencing through a pricing algorithm, \textit{Smart Pricing (SP)}.
Airbnb is a platform that enables users to rent out unused homes or vacant rooms to travelers. Providers of these accommodations, referred to as hosts, are charged a 3\% service fee on their earnings by Airbnb.\footnote{As noted on Airbnb's website: \url{https://www.airbnb.jp/help/article/1857}, accessed on May 23, 2024.} Since 2015, Airbnb has offered a pricing algorithm, SP, which allows hosts to delegate price determination to the algorithm.\footnote{\url{https://www.cnbc.com/2015/11/12/airbnb-launches-smart-pricing-for-hosts.html}, accessed on April 12, 2024.} Hosts retain the option to use SP at their discretion.

\begin{figure}[H]
    \centering
    \includegraphics[width=\linewidth]{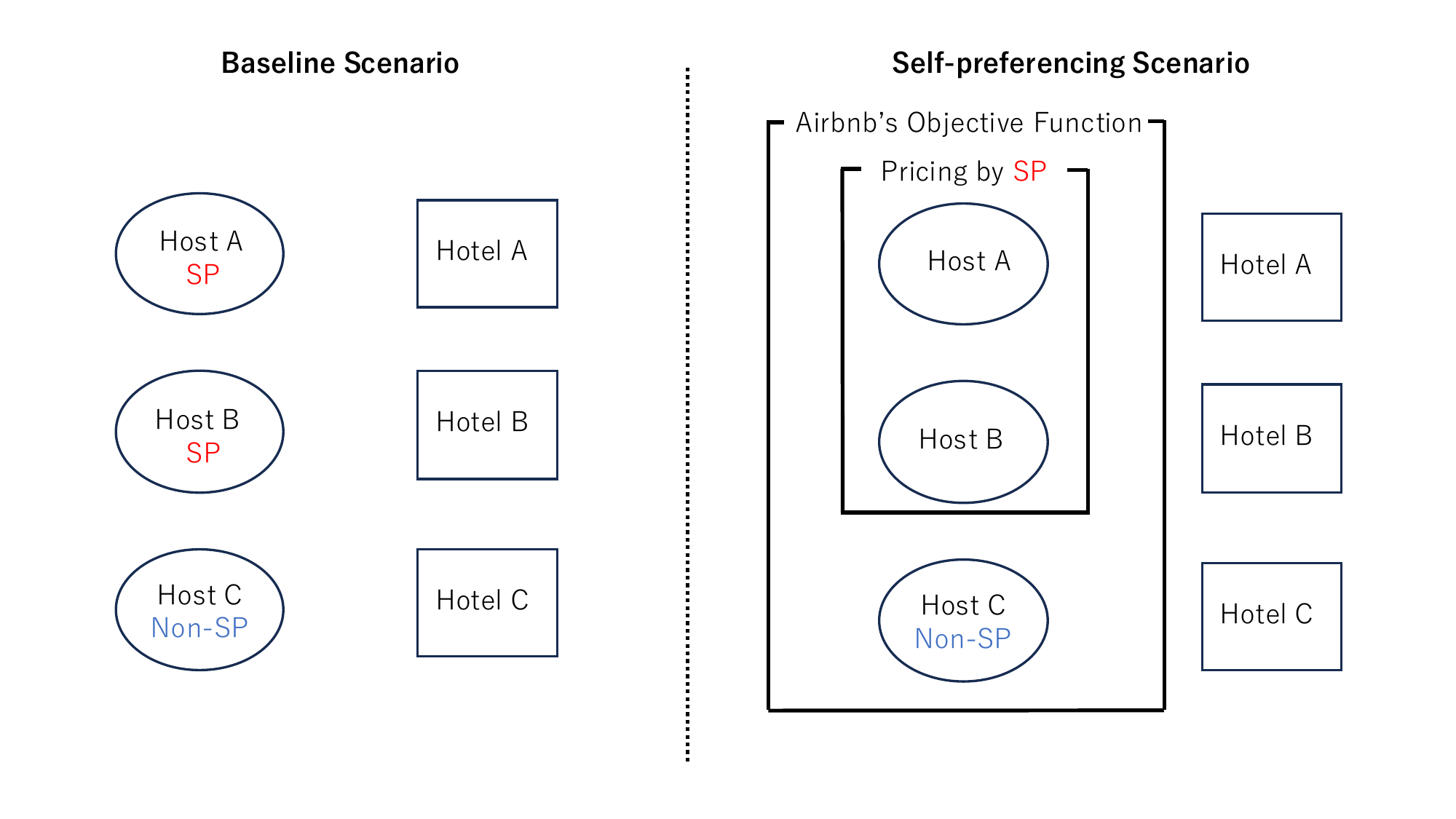}
    \caption{Market Structure Under Each Scenario}
    \label{fig:market_structure}
    \begin{flushleft}
        \small Note: This figure shows an example of the market structure in each scenario, with hosts A and B using SP. In the baseline scenario, each host on Airbnb also competes on price with each other. In the self-preferencing scenario, Airbnb determines the price of goods for each host using the SP, leading to a reduction in the intensity of price competition compared to the baseline. Airbnb solves the profit maximization problem by also considering the commission earned from Host C's revenue.
    \end{flushleft}
\end{figure}

In this paper, I define Airbnb's self-preferencing as the practice of using SP to maximize the platform's total commission revenue, rather than prioritizing individual host revenue.\footnote{Since Airbnb cannot observe the marginal cost of individual hosts, it maximizes revenue rather than profit.} This approach aligns with Airbnb's short-run incentives, as maximizing commission revenue across the platform often conflicts with optimizing revenue for individual hosts. 
To analyze this behavior, I consider two distinct scenarios:
\begin{enumerate}
    \item \textbf{Baseline scenario.} Airbnb adjusts SP to maximize the revenue of a host using SP.
    \item \textbf{Self-preferencing scenario (counterfactual).} Airbnb adjusts SP to maximize its total commission revenue.
\end{enumerate}
Figure \ref{fig:market_structure} graphically provides an example of market structures in each scenario. I assume the self-preferencing scenario to be counterfactual because Airbnb claims that SP is designed to prioritize host revenue \citep{airbnb2018customized}. By comparing welfare outcomes across these scenarios, I empirically evaluate the social welfare effects of Airbnb's self-preferencing.

I constructed unique data by scraping accommodation information from the websites of Airbnb and Booking.com.
The dataset covers listings from Tokyo's 23 wards, collected between May 12 and July 13, 2024. Key variables include accommodation prices, the number of reviews, average ratings for various criteria (such as cleanliness, location quality, and staff communication), and the number of available rooms for hotels and private accommodations.


Using this data, I estimate a static Bertrand competition model between Airbnb hosts and hotels listed on Booking.com. The analysis proceeds in three steps. First, the consumer demand is modeled using the random coefficient nested logit demand model, estimated with the \cite{blp1995}'s algorithm (BLP). This model incorporates a nested structure based on price ranges to capture that the substitution patterns between Airbnb accommodations and hotels weaken when their prices differ significantly, as found by \cite{zervas2017rise}. Second, assuming Bertrand competition with differentiated products, the marginal costs of hosts and hotels are estimated by first-order conditions. In the counterfactual scenario representing self-preferencing, Airbnb adjusts host prices using SP to maximize total platform-wide commission revenue. This adjustment also accounts for the revenues of hosts not using SP, thereby relaxing price competition. Finally, I compute the differences in social welfare between the baseline and self-preferencing scenarios, quantifying the empirical impact of Airbnb’s self-preferencing on social welfare within the framework of short-run price competition.

The results indicate that Airbnb's self-preferencing significantly negatively impacts social welfare.
Specifically, when Airbnb engages in self-preferencing through SP, social welfare decreases by an average of 5.08\% per market. This decline is driven largely by a notable increase in prices set by hosts utilizing SP, with prices rising by up to 1.5 times compared to the baseline scenario.
This translates to an average daily welfare loss of 115,400 yen per market. 
Aggregating this estimate across Tokyo’s 23 wards yields a total daily welfare loss of 2.65 million yen.
On an annual basis, this amounts to a welfare reduction equivalent to approximately 14.90\% of Tokyo's 2023 vacation rental market.
A major driver of these welfare losses is the substantial increase in Airbnb’s total commission revenue, which rose by an average of 37.73\%.
This highlights Airbnb's strong incentive for engaging in self-preferencing, as the platform captures a greater share of market surplus through its pricing algorithms.
These results demonstrate how self-preferencing by platforms can harm social welfare by prioritizing platform interests over broader market efficiency. Given the lack of transparency requirements for algorithms, the findings call for urgent discussions on competition policy and algorithmic accountability to mitigate such adverse effects.


\noindent
\textbf{Related Literature.} This paper contributes to three key strands of literature: (i) platform self-preferencing, (ii) algorithmic pricing, and (iii) the impact of Airbnb on the hotel industry.

First, regarding the platform self-preferencing, a growing body of literature has examined practices such as manipulating search results \citep{de2019model, lam2021platform, kittaka2022dual}, altering recommendation systems and launching private brand products on platforms like Amazon \citep{zhu2018competing, wen2019threat, Chen_and_Tsai_2024}.\footnote{See \cite{kittaka2023self} for a literature review of self-preferencing by platforms.} However, empirical analyses investigating the impact of self-preferencing on social welfare have yet to be developed. This is largely because the algorithms used by platforms are unobservable, making it difficult to infer causal relationships between social welfare and self-preferencing. This paper contributes to this literature by constructing a Bertrand competition model between the platform and non-platform firms, quantifying the welfare impact under the self-preferencing scenario.

Next, in the literature on pricing algorithms, prior studies have explored simulations of price competition using machine learning \citep{calvano2020artificial, klein2021autonomous}, theoretical studies on collusion strategies \citep{salcedo2015, lamba2022pricing, Asker-et-al-2022-AEA, Brown-and-MacKay2023}, and empirical studies of the impact of pricing algorithms on prices \citep{assad2024algorithmic, musolff2024algorithmic, hortaccsu2024organizational}. Among these, this paper is particularly related to \citet{harrington2022effect}.\footnote{See also \cite{johnson2023platform}, which examines how Amazon can design its recommendation system to prevent collusion when sellers use pricing algorithms and provide the simulation result using machine learning.} He theoretically analyzed the incentives of competing firms in a duopoly market for differentiated products and a firm that develops and sells pricing algorithms. 
My analysis differs from \citet{harrington2022effect} by focusing on a platform that provides pricing algorithms to its users for free. 
Airbnb has an incentive to maximize its commission revenue from transactions through pricing algorithms.\footnote{Airbnb's commission fee is fixed at $3\%$, a rate established by Airbnb and unchanged from 2015 through 2024. For studies that model the fees set by platforms selling their products and analyze their welfare effects, see for example \cite{anderson2021hybrid, etro2023hybrid}.}
By measuring the effect of this platform-specific incentive on social welfare, I make a novel contribution to the literature. However, it should be noted that I do not analyze the positive welfare effects of pricing algorithms, such as improved demand forecasting or reduced pricing costs, as discussed by \cite{miklos2019collusion} and \cite{Connor-Wilson-2021reduced}.

Finally, this paper contributes to the literature on the impact of private accommodations on the hotel industry and consumers. 
Numerous studies have analyzed Airbnb's effects on incumbent hotels \citep{horn2017home, zervas2017rise, dogru2020effects, barron2021effect, farronato2022welfare}. Notably, \cite{farronato2022welfare} modeled competition between Airbnb hosts and hotels as Cournot competition, empirically assessing the effects of Airbnb's entry on hotel revenues, prices, and social welfare. However, their analysis predates the introduction of Airbnb's SP and thus does not consider the impact of self-preferencing through SP.
This paper diverges by focusing on short-term incentives and assuming Bertrand competition, offering a distinct perspective from \cite{farronato2022welfare}. 

I have organized the rest of the paper as follows. Section \ref{Accommodation_Industry_and_Airbnb} provides an overview of the background of the accommodation industry, while Section \ref{Data} describes the data. Section \ref{Model_and_Estimation_Strategy} presents the price competition model between Airbnb hosts and hotels. Section \ref{Estimation_Results} discusses the estimation results of model primitives. Section \ref{counterfactual_simulation_results} details the results of the counterfactual simulation, including changes in equilibrium prices and Airbnb's commission revenue when transitioning from the baseline scenario to the self-preferencing scenario, as well as the welfare effects of self-preferencing.
Finally, Section \ref{Conclusion} concludes.

\section{Industry and Data}
This section explains why Airbnb and the accommodation industry are particularly suitable for the study of the welfare effects of self-preferencing and describes its key features to motivate my structural model in the next section.

\subsection{Accommodation Industry and Airbnb}
\label{Accommodation_Industry_and_Airbnb}
\begin{figure}[H]
        \centering
        \includegraphics[width=0.7\linewidth]{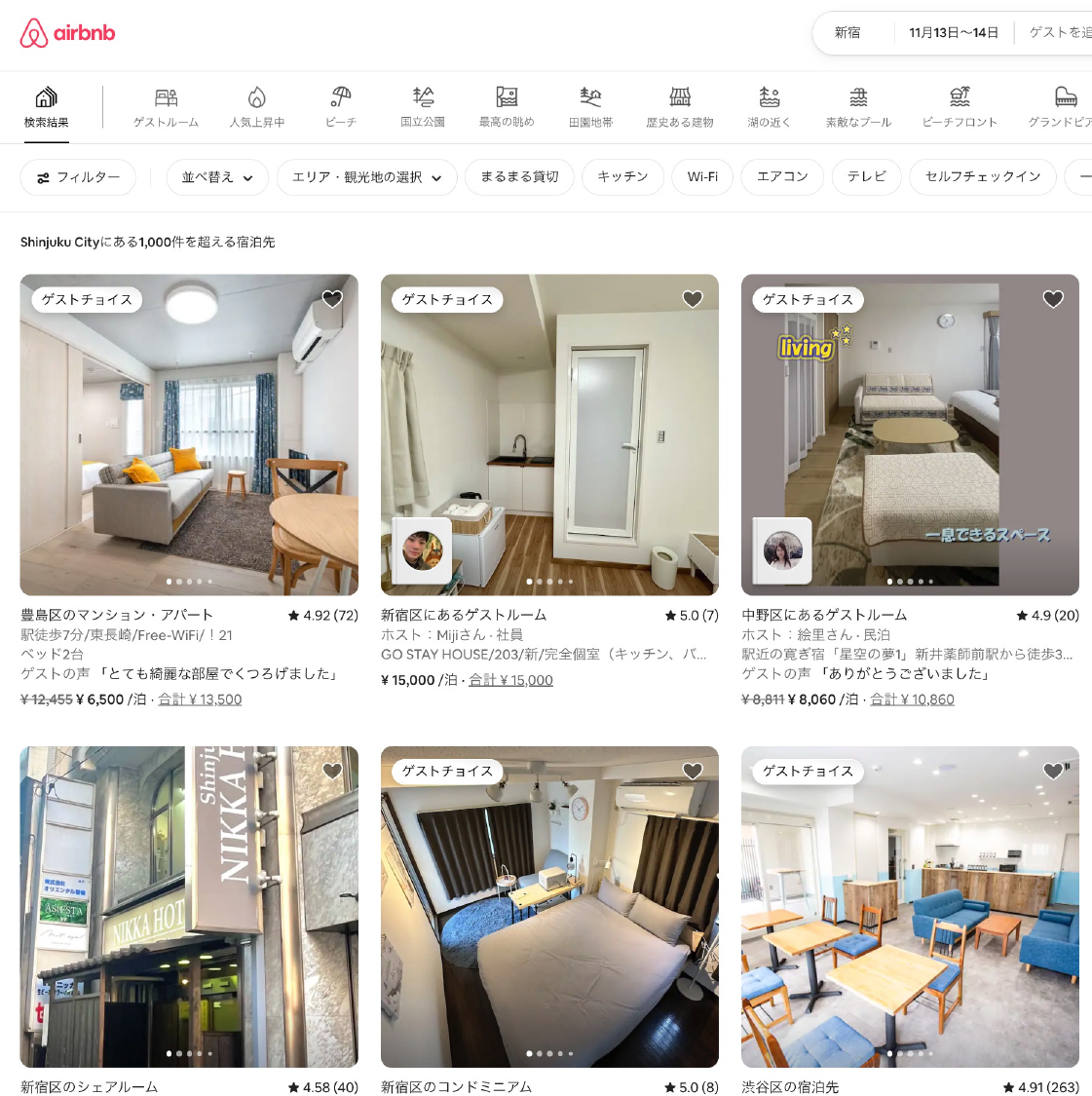}
        \caption{The example of Airbnb search result (retrieved on November 1, 2024)}
        \label{fig:example_of_airbnb_search_result}
\end{figure}

\begin{figure}[H]
    \centering
    \includegraphics[width=0.8\linewidth]{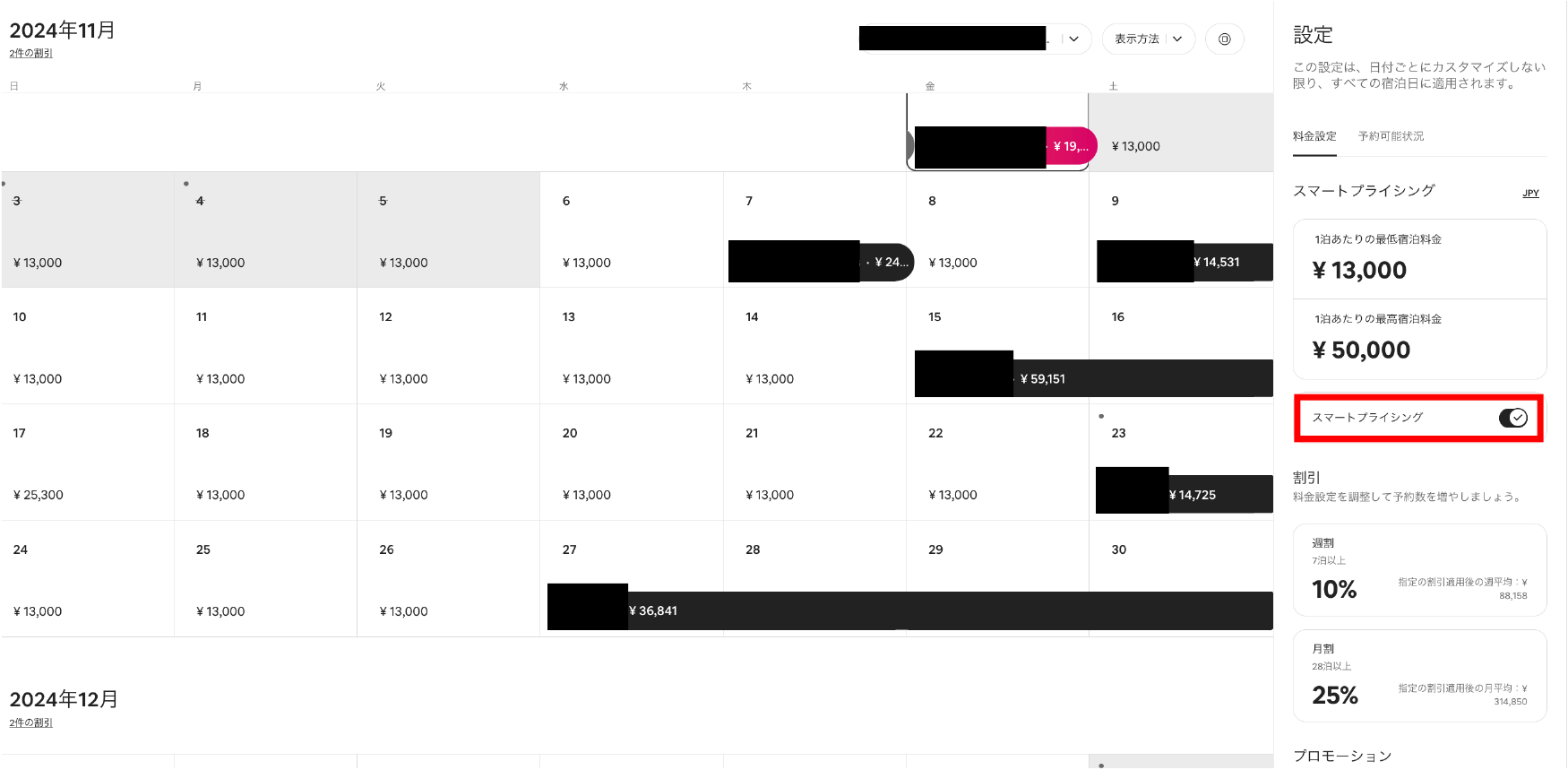}
    \caption{The screenshot of the host's calendar interface (where Smart Pricing can be configured)}
    \label{fig:screenshot_of_the_hosts_calendar_with_SP}
    \begin{flushleft}
    \small Note: Each host determines prices through this calendar interface. By enabling ``Smart Pricing" (highlighted in a red box) and setting the minimum and maximum nightly rates, SP automatically suggests prices. The names and icons of booked guests and hosts have been redacted for privacy. The screenshot was captured on November 1, 2024.
    \end{flushleft}
\end{figure}

Airbnb is an online platform offering accommodations and experiences, featuring over 5 million hosts and properties worldwide. Since its launch in 2007, Airbnb has facilitated more than 2 billion guest stays. Figure \ref{fig:example_of_airbnb_search_result} and Figure \ref{fig:screenshot_of_the_hosts_calendar_with_SP} illustrate key aspects of the platform, with the former showcasing a sample Airbnb search result and the latter providing a screenshot of a host's pricing calendar featuring SP.
In 2015, Airbnb introduced SP, which they claim optimizes prices to maximize individual host revenue \citep{airbnb2018customized}. 
I adopt the scenario where Airbnb uses SP to maximize host revenue as the baseline, conducting counterfactual simulations under a self-preferencing scenario to evaluate its impact.
In Japan, other vacation rental intermediaries, such as \textit{STAY JAPAN} and \textit{Airtrip Minpaku}, operate in the market. However, Airbnb dominates the industry. As of April 12, 2024, the number of rooms available for reservation in Tokyo on May 12 was as follows: Airbnb listed 1,243 rooms, STAY JAPAN listed 16, and Airtrip Minpaku listed 49. Given (i) Airbnb’s provision of SP as a key platform feature and (ii) its dominant position in Tokyo’s vacation rental market, Airbnb serves as an ideal subject for examining the effects of platform-driven self-preferencing on social welfare.

Empirical studies have shown that Airbnb's entry into the accommodation industry reduces hotel revenues and prices, suggesting a substitutive relationship between the two \citep{zervas2017rise, farronato2022welfare}. However, there is no evidence of competition between hotels and hosts in Tokyo. To guarantee the robustness of results, in Section \ref{Estimation_Results}, I estimate models under two contrasting assumptions: one where Airbnb hosts and hotels are in competition and engage in price competition, and another where they do not compete. The latter assumption follows the framework adopted by \cite{ForoughifarMehta2024}. This approach serves as a robustness check for the results, ensuring that the findings are not sensitive to the assumed nature of competition within the accommodation industry.

\subsection{Data}
\label{Data}
I have constructed unique data for Airbnb and hotel listings on Booking.com by scraping web data from their websites. The dataset covers listings within Tokyo’s 23 wards, from May 12 to July 13, 2024.
I observe the daily prices, vacancies, reviews per room, and average review ratings for various criteria, including cleanliness, communication with staff, location quality, and overall room evaluation. 
Since Airbnb hosts and hotels adopt dynamic pricing, adjusting their prices in response to demand fluctuations as the check-in date approaches, I collected all data one month before the check-in date to minimize the influence of dynamic pricing.\footnote{\cite{pan2021costly} and \cite{huang2024seller} study how Airbnb hosts set prices to respond to temporal demand shocks and how the check-in date prices vary depending on the time gap between the check-in date and the booking date.}
In Appendix \ref{data_explanation}, I present the reason why I chose Booking.com as a source of hotel information, as well as a detailed explanation of the data.

\subsubsection{Descriptive Statistics}
I display descriptive statistics in Table \ref{Descriptive_Statistics}. This table presents the means and standard deviations of prices, the number of reviews, evaluation indices (such as location, staff communication, cleanliness, and overall room rating), and the number of available units, separately for Airbnb accommodations and hotels listed on Booking.com across all periods in the 23 wards of Tokyo. 

\begin{figure}[H]
    \centering
    \includegraphics[width=0.8\linewidth]{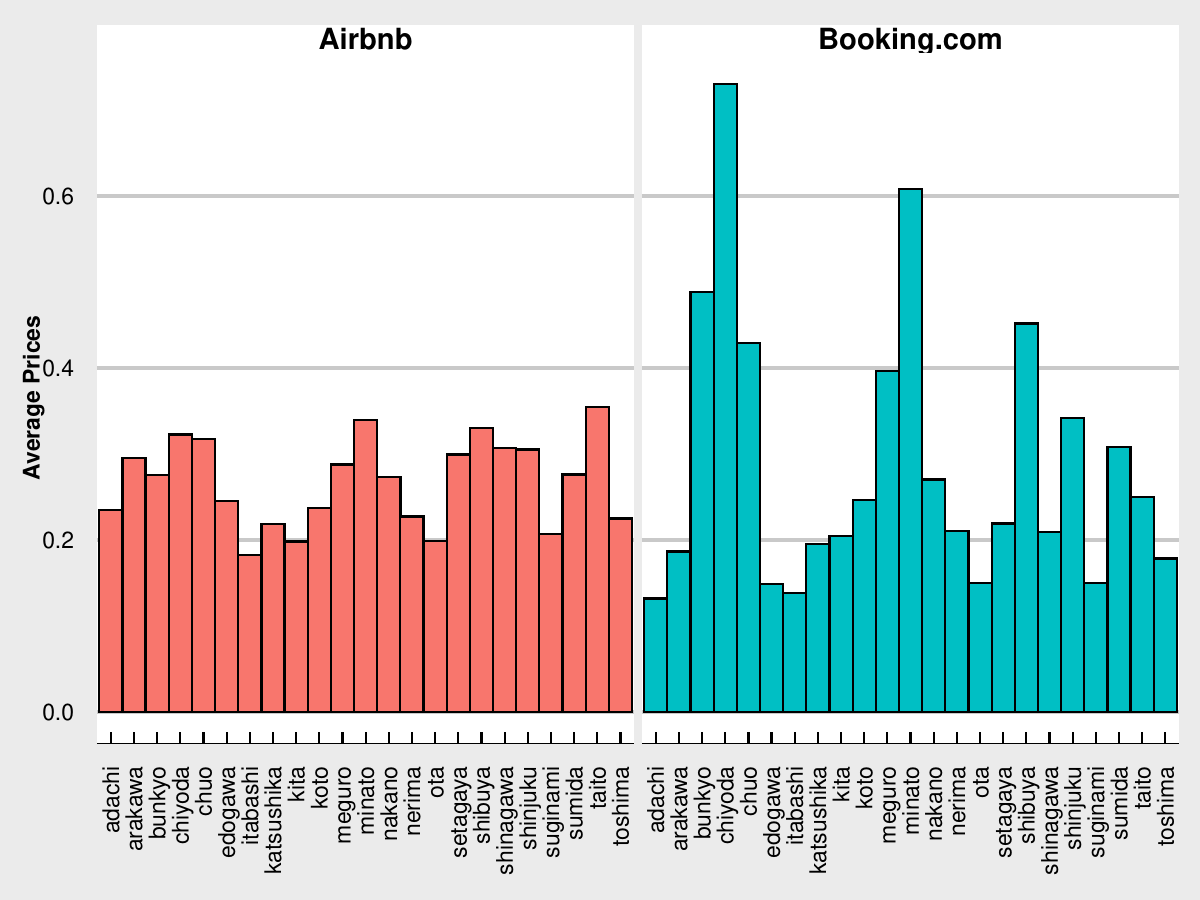}
    \caption{Average Prices by City}
    \label{fig:ave_price_by_city}
\end{figure}

These descriptive statistics reveal two key heterogeneities that underscore the differences between Airbnb hosts and hotels: prices and ratings. On average, hotels tend to have higher prices compared to Airbnb hosts (0.337 for hotels and 0.287 for Airbnb hosts, with prices normalized by dividing by 100,000). This disparity is largely driven by the presence of luxury hotels. Figure \ref{fig:ave_price_by_city} compares the average prices of Airbnb hosts and hotels across Tokyo’s 23 wards. The overall average price is elevated by the concentration of luxury hotels in areas like Chiyoda and Minato wards. However, in other areas, Airbnb hosts generally exhibit higher average prices than hotels.
This observation motivates me to include a dummy variable in the demand model to distinguish whether a listing is operated by an Airbnb host or a hotel.

The average ratings for evaluation indices are statistically significantly lower for hotels than for Airbnb accommodations, while the number of reviews is higher for hotels. The following factor can explain this: differences in incentives within the evaluation systems may also play a significant role. As discussed by \cite{belleflamme2021economics}, both Airbnb hosts and hotels derive positive gains from favorable reviews and ratings, creating incentives for strategic distortions of them. The data I obtained also suggest the possibility of bias due to such “fake reviews” \citep{xu2015commerce, zervas2021first, he2022market}. \cite{he2022market} examines how fake reviews are used on Amazon Marketplace, where reviewers sell their services in private Facebook groups and receive payments through PayPal. Their findings indicate that the average ratings of products decline once firms cease purchasing fake reviews, suggesting that such reviews are predominantly used to promote low-quality products. Similarly, complementary work by \cite{zervas2021first} reveals that nearly 95\% of Airbnb properties worldwide hold an average rating of 4.5 or 5 stars, with properties rated below 3.5 stars being virtually nonexistent. In contrast, hotels on TripAdvisor exhibit significantly lower average ratings of 4.1 stars with greater variation between reviews. While vacation rentals on TripAdvisor achieve relatively high ratings, only about 85\% of these properties have average ratings of 4.5 stars or higher. These observations are highly consistent with the rating patterns in my data.

\subsubsection{Criteria for Determining the Use of Smart Pricing}
Whether Airbnb hosts are using SP cannot be directly observed from my data. I therefore determine that a host is using SP if they meet the following two criteria:\\
\noindent
\textbf{Criterion 1}: “Operating Low-Priced Listings”.
Hosts can significantly minimize their price-setting costs.\footnote{\citet{zbaracki2004managerial} empirically analyzed pricing costs using firm data. They identified that pricing costs, considering menu costs, information gathering, decision-making, and internal communication costs, account for 1.22\% of revenue. Hosts operating low-priced listings, which have a lower degree of product differentiation, are expected to have stronger incentives to reduce these costs.}
Thus, hosts operating low-priced listings, which are less differentiated, are more likely to adopt SP. \citet{zhang2021frontiers} used Airbnb data from the United States between 2015 and 2017 and showed that the average monthly prices for hosts using SP and those not using SP were $\$179.80$ and $\$194.61$, respectively.\footnote{The number of hosts using SP was 2,118, while the number not using SP was 7,278. While \citet{zhang2021frontiers} obtained information on SP usage directly from Airbnb, the data used in this study does not allow for direct observation of whether hosts use SP.}
The relatively lower prices among hosts using SP support the validity of Criterion 1. The classification of whether a host operates low-priced listings is conducted as follows: (i) perform 5-level clustering based on prices for both Airbnb hosts and hotels on Booking.com using the k-means method, (ii) interpret listings in the bottom two clusters as low-priced.
The clustering results are shown in Table \ref{K-means}.

\noindent
\textbf{Criterion 2}: “Engaging in Dynamic Pricing”.
Dynamic pricing refers to adjusting prices in response to changes in demand. In the accommodation industry, sellers operate with fixed capacities, which typically leads to lower prices as the check-in date approaches. Under Smart Pricing, prices are updated daily, so hosts who change their prices daily are considered to meet Criterion 2.
As a result, the proportion of Airbnb hosts utilizing dynamic pricing was found to be 52.13\%.

Following the above procedure, I found that 16.7\% of Airbnb hosts met both criteria. This is close to the figure reported in previous studies using datasets with known SP usage \citep{zhang2021frontiers, ForoughifarMehta2024}, where the proportion was 22\% in both datasets, indicating consistency with past findings. Table \ref{SP_ratio} shows how many listings use SP for each cluster, restricting the sample to Airbnb only.

\begin{figure}[H]
    \centering
    \includegraphics[width=0.8\linewidth]{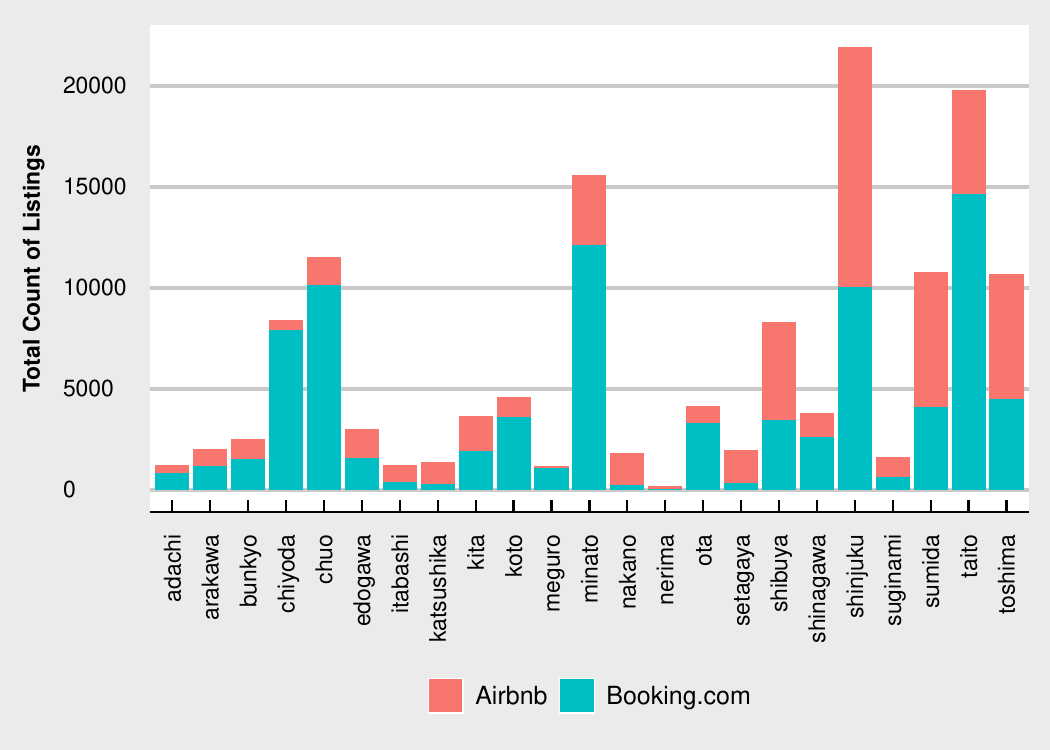}
    \caption{Total Number of Airbnb and Hotel Listings by City}
    \label{fig-Airbnb-ratio}
    \begin{flushleft}
        \small Note: The vertical axis is a total count of how many listings exist in each city.
    \end{flushleft}
\end{figure}

\begin{figure}[H]
    \centering
    \begin{minipage}{0.49\linewidth}
        \centering
        \includegraphics[width=\linewidth]{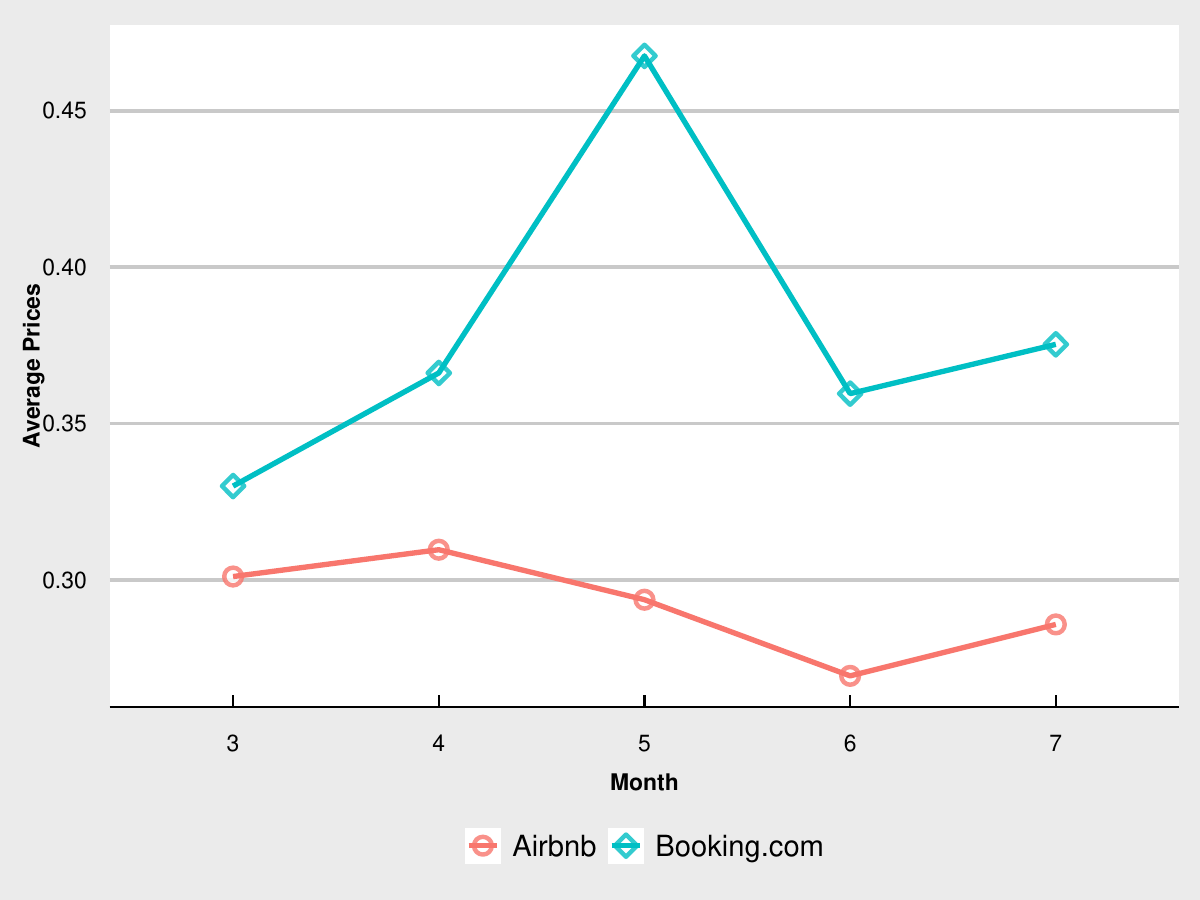}
        \caption*{(a) Average Prices by Month}
        \label{fig:ave_price_by_month}
    \end{minipage}
    \hfill
    \begin{minipage}{0.49\linewidth}
        \centering
        \includegraphics[width=\linewidth]{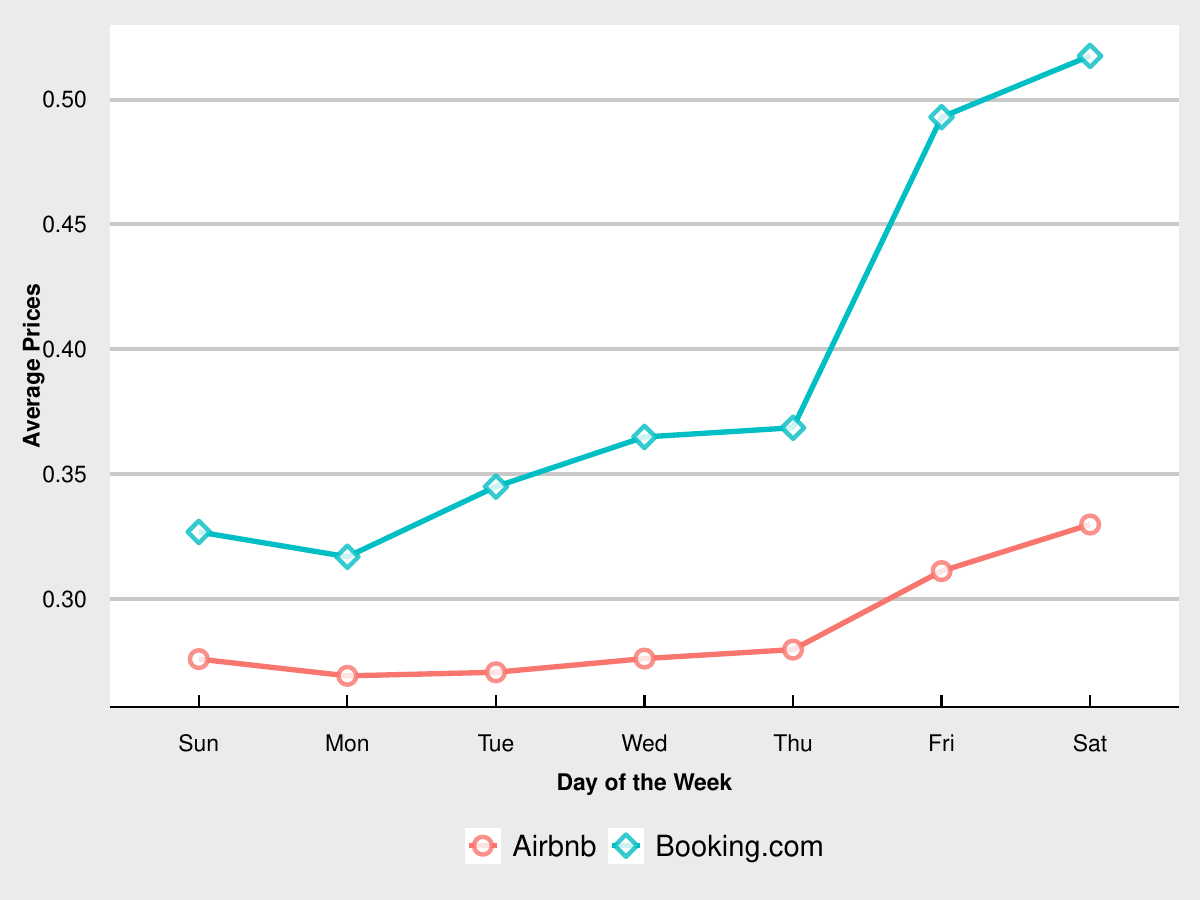}
        \caption*{(b) Average Prices by Day of Week}
        \label{fig:ave_price_by_day_of_week}
    \end{minipage}
    \caption{Average prices: (a) by month and (b) by day of the week.}
    \label{fig:ave_price_side_by_side}
\end{figure}

I observed notable variations among price clusters in Table \ref{K-means}, which suggest the need to incorporate a nested structure into the Bertrand competition model. Similarly, I found significant differences in accommodation prices and supply across months, wards, and days of the week. Figure \ref{fig-Airbnb-ratio} illustrates the total number of Airbnb listings and hotels in each ward, revealing substantial disparities in supply. These variations likely arise from factors such as (i) differences in demand driven by the presence of tourist attractions, and (ii) variations in supply costs linked to rent differences across wards. To address (i), my model incorporates fixed effects for each of the 23 wards, capturing the heterogeneity in market characteristics. Figure \ref{fig:ave_price_by_city} and Figure \ref{fig:ave_price_side_by_side} further highlight differences in average prices by ward, month, and day of the week, reinforcing the need for fixed effects in the consumer demand model.
In contrast, addressing (ii) within the supply side is infeasible due to the lack of observable variables that capture daily fluctuations in supply factors across the 23 wards of Tokyo. For instance, rent for accommodations could serve as a supply factor. However, since rents do not fluctuate daily, they cannot be incorporated into the supply-side model, which defines markets based on a pair of daily and area.
\section{Model and Estimation Strategy}
\label{Model_and_Estimation_Strategy}
In this section, I describe consumer demand and the short-run price competition model that I use to estimate the welfare effects of self-preferencing and explain the estimation strategy.

\subsection{Consumer Demand}
\label{Consumer_Demand_Model}
I use the random coefficient nested logit (RCNL) model to estimate consumer demand. The RCNL model has been applied in several recent empirical articles \citep{ciliberto2014does, miller2017understanding, Igami2024}.

I define a market $m$ by a pair of days and one of Tokyo's 23 wards. Each consumer purchases one of the observed products ($j = 1,\cdots, J_m$) or selects the outside option ($j=0$).
Consumer $i\in\{1,\ldots, I_m\}$'s indirect utility from product $j$ in market $m$ is
\begin{equation}
\label{indirect-utility-specification}
    u_{ijm} = (\alpha + \Sigma {v}_{im}) p_{jm} + \mathbf{X}'_{jm}\boldsymbol{\beta} + \xi_{jm} + \bar{\varepsilon}_{ijm}
\end{equation}
where $\mathbf{X}$ is a vector of observable product characteristics, ${v}_{im}$ is i.i.d normal distribution $N(0,1)$, the parameter $\Sigma$ govern the extent of variation in tastes for $p_{jm}$ across consumers with different taste shocks $v_{im}$,  $\xi_{jm}$ represents unobserved product quality. I incorporate city, month, and the day of the week fixed effects\footnote{I cannot incorporate the product fixed effect because of the curse of dimensionality (see \cite{PyBLP} for details).} for $\xi_{jm}$ so that $\xi_{jm} = \xi_{\text{city}} + \xi_{\text{month}} + \xi_{\text{day of week}} + \Delta\xi_{jm}$. I control for each fixed effect using dummy variables and $\Delta\xi_{jm}$ is left as a structural error. $\bar{\varepsilon}_{ijm}$ is a stochastic term.
The observable product characteristics $\mathbf{X}$ include the average review scores for various criteria (cleanliness, communication with staff, location quality, and the overall evaluation) and an indicator for whether the product is on Airbnb. 

I assume that $\bar{\varepsilon}_{ijm}$ can be decomposed by
\begin{equation}
    \bar{\varepsilon}_{ijm} = \zeta_{igm} + (1-\rho)\varepsilon_{ijm}
\end{equation}
where $g$ represents the pre-determined group which I define a pair of price clusters (as shown in Table \ref{K-means}) and whether the listing is Airbnb or hotels, $\varepsilon_{ijm}$ is the i.i.d. extreme value, $\zeta_{igm}$ has the unique distribution such that $\bar{\varepsilon}_{ijm}$ is extreme value, and $\rho\in[0,1)$ is a nesting parameter. Larger values of $\rho$ correspond to a greater correlation in preferences for products of the same group. Finally, we normalize the indirect utility of the outside good such that $u_{i0m} = \varepsilon_{ijm}$. To sum up, the parameters we need to estimate are $(\alpha, \boldsymbol{\beta}, \Sigma, \rho)$. In Appendix \ref{appendix_demand_model_and_welfare_measure}, I present the equations for the market share and demand of product $j$, as well as the formulas for calculating consumer surplus, producer surplus, and social welfare.

This RCLN model specification was motivated by key features of my empirical context and dataset. As shown in Table \ref{K-means}, significant heterogeneity exists between Airbnb and hotels across different price clusters, which is captured by a nested structure and the random coefficient $\Sigma$ to account for substitution patterns flexibly.\footnote{I explored additional random coefficients; however, the estimates produced were counter-intuitive due to perhaps multi-collinearity (see \cite{grigolon2014nested} for the detail).} It is also natural to assume that consumers exhibit common preferences regarding whether the accommodation is listed on Airbnb or is a hotel, and this is captured using a dummy variable for Airbnb.

\subsection{Estimation Strategy}
I estimate the demand model by BLP algorithm using \cite{PyBLP}'s Python package \textit{PyBLP}.\footnote{I thank Jeff Gortmaker for sharing these details especially how to implement the specific counterfactual simulation.} Prices and within-nest market shares are likely to be correlated with the structural error term $\Delta\xi_{jm}$ because firms set prices with knowledge of product- and market-specific consumer valuations. This creates a standard endogeneity problem. To address this, I use three types of instrumental variables: (i) the number of products in each category defined by K-Means clustering, (ii) the measures of product differentiation (``differentiation instruments”) proposed by \cite{GandhiHoude2023} which intuitively capture the relative isolation of each product in characteristics space, and (iii) own product characteristics, $\mathbf{X}_{jm}$.\footnote{i.e., I assume $\mathbb{E}[\Delta\xi_{jm}\mid \mathbf{X}_{jm}] = 0$. This assumption that the product characteristics of each good are uncorrelated with the structural error is reasonable since the focus is on short-term decision-making.} For differentiation IVs, I employ both the Euclidean distances across continuous characteristics between a focal product and others and their interactions. The interaction terms capture the covariance between two dimensions of differentiation.

Following \cite{farronato2022welfare}, I exclude products with a market share of less than $0.5\%$ from the analysis. This exclusion is necessary because the presence of products with market shares approaching zero complicates estimation and prevents the BLP algorithm from converging. Regarding the potential market size $I_m$ in market $m$, I adopt the approach from the literature on structural estimation using Airbnb \citep{ li2019competitive, farronato2022welfare, ForoughifarMehta2024}, defining $I_m \equiv \text{``number of rooms in each market} \times 2$”. Table \ref{Descriptive_statistics_for_estimation} presents the descriptive statistics for the dataset after applying the aforementioned sample restrictions. The characteristics observed are largely consistent with those of the full sample, as shown in Table \ref{Descriptive_Statistics}. Furthermore, Table \ref{The_Number_of_Firms_in_Each_Market} provides the number of hosts using SP, those not using it, and the number of hotels in each market after applying the sample restrictions, where a market is defined as the pair of a specific day and one of Tokyo's 23 wards. On average, there are 7.28 firms per market, highlighting the oligopolistic environment of these markets. 

\subsection{Supply-side}
I consider SP hosts, non-SP hosts, hotels, and Airbnb as the game players. That is, the set of players and products are $\mathcal{F}_m = \mathcal{F}_{\text{SP}, m} \cup \mathcal{F}_{\text{non-SP}, m} \cup \mathcal{F}_{\text{hotel}, m} \cup \qty{\text{Airbnb}}$ and $\mathcal{J}_m = \mathcal{J}_{\text{SP}, m} \cup \mathcal{J}_{\text{non-SP}, m} \cup \mathcal{J}_{\text{hotel}, m}$, respectively.
Note that SP hosts do not determine their prices and I assume that this decision was made by hosts exogenously.
In the market $m$, each player $f\in \mathcal{F}_m\setminus\mathcal{F}_{\text{SP},m}$ decides simultaneously the price $p_{jm}\in [0,\infty)$ for product $j\in \mathcal{J}_{f,m}$. 
They have the complete information. Each host and hotel receives the profits which are realized at pure-strategy Nash equilibrium. Airbnb obtains the commission revenue from hosts' revenue:
\begin{equation*}
    \Pi_{\text{Airbnb}, m} = \tau\qty(\sum_{j \in \mathcal{J}_{\text{Airbnb}, m}} p_jq_j)
\end{equation*}
where $\tau$ denotes the commission rate charged by Airbnb to hosts ($\tau = 0.03$) and $\mathcal{J}_{\text{Airbnb}, m} = \mathcal{J}_{\text{SP}, m}\cup\mathcal{J}_{\text{non-SP}, m}$ represents the set of all hosts' products on Airbnb.
I will omit the market index $m$ for simplicity in the following description.

\subsubsection{Baseline Scenario}
In the baseline scenario, Airbnb adjusts SP to maximize the revenue of an individual host using SP, while each non-SP host and hotel $f \in \mathcal{F}_{\text{non-SP}} \cup \mathcal{F}_{\text{hotel}}$ solves its profit maximization problem. 
Therefore, we can interpret the objective function of each player $f \in \mathcal{F} \setminus \qty{\text{Airbnb}}$ as follows:
\begin{equation}
    \Pi_{f} = \sum_{j\in \mathcal{J}_f}\qty(p_{j}-\mathbf{1}(j)\cdot\text{mc}_{j})q_{j}
\end{equation}
where $\text{mc}_{j}$ is the marginal cost, assumed to be constant, and $\mathbf{1}(\cdot)$ is an indicator function that equals 1 if $j \notin \mathcal{J}_{\text{SP}}$, and 0 if $j\in\mathcal{J}_{SP}$. By first-order conditions, we have
\begin{equation}
\label{FOC}
    q_{j} + \sum_{k\in\mathcal{J}_f}(p_k-\mathbf{1}(k)\cdot\text{mc}_{k})\frac{\partial q_k}{\partial p_{j}} = 0\quad \forall f\in \mathcal{F},~\forall j\in \mathcal{J}_f.
\end{equation}
I introduce matrices and vectors as follows:
\begin{equation}
    \mathbf{q} = \begin{bmatrix}
        q_{1}\\
        \vdots\\
        q_{J}
    \end{bmatrix},~ \mathbf{p} = \begin{bmatrix}
        p_{1}\\
        \vdots\\
        p_{J}
    \end{bmatrix},~\frac{\partial \mathbf{q}}{\partial \mathbf{p}} = \begin{bmatrix}
        \frac{\partial q_1}{\partial p_1} & \cdots & \frac{\partial q_{J}}{\partial p_1}\\
        \vdots & \ddots & \vdots\\
        \frac{\partial q_{1}}{\partial p_{J}} & \cdots & \frac{\partial q_{J}}{\partial p_{J}} 
    \end{bmatrix},~\mathbf{mc} = \begin{bmatrix}
        \mathbf{1}(1)\cdot\text{mc}_1\\
        \vdots \\
        \mathbf{1}(J)\cdot\text{mc}_J
    \end{bmatrix}.
\end{equation}
I define a $J \times J$ ownership matrix $\mathbf{H}$, whose $(i, j)$ element is given by
\begin{equation}
    {H}_{ij} = \begin{cases}
        1 & \text{if }(i,j)\subset \mathcal{J}_f\\
        0 & \text{otherwise}.
    \end{cases}
\end{equation}
Using this, the first-order conditions \eqref{FOC} can be rewritten as follows:
\begin{align}
    \mathbf{q} + \mathbf{H}\odot \frac{\partial \mathbf{q}}{\partial \mathbf{p}}[\mathbf{p} - \mathbf{mc}] &= \mathbf{0}\\
     \mathbf{mc} &= \mathbf{p}-(\Delta \mathbf{D})^{-1}\mathbf{q} \label{mc}
\end{align}
where $\odot$ denotes the Hadamard product. $\Delta \mathbf{D}$ is defined as
\begin{equation}
    \Delta \mathbf{D} \equiv -\mathbf{H}\odot \frac{\partial \mathbf{q}}{\partial \mathbf{p}}
\end{equation}
and is assumed to be a nonsingular matrix. Since marginal costs are assumed to be constant, the marginal cost for each non-SP host and hotel can be calculated from equation \eqref{mc}. Therefore, in the case of the baseline scenario, the observed prices are substituted into equation \eqref{mc} to compute marginal costs.

\subsubsection{Self-preferencing Scenario}
In the self-preferencing scenario, Airbnb adjusts SP to maximize its total commission revenue.
In Table \ref{scenario2-game-environment}, I present the strategic environment under this scenario.
In particular, what differs from the baseline scenario is the objective function of Airbnb's maximization problem:
\begin{equation*}                       
    \max\limits_{p_j}\underbrace{\tau\qty(\sum\limits_{j\in{\mathcal{J}_{\text{Airbnb}}}}~p_jq_j)}_{\text{Platform-wide Commission Revenue}}~ \forall j \in \mathcal{J}_{\text{SP}}.
\end{equation*}
That is, Airbnb's objective function is the platform-wide commission revenue rather than individual SP-host revenue.
Since we have already estimated demand and calculated the marginal cost, the counterfactual equilibrium prices can be calculated from the new first-order conditions under the environment described in Table \ref{scenario2-game-environment}, using the contraction mapping algorithm proposed by \cite{morrow2011fixed}.


\section{Estimation Results}
\label{Estimation_Results}
\subsection{Consumer Demand}
Table \ref{Demand_Estimates} presents the results of the demand estimation. Columns (i) and (iii) reflect the main RCNL specification, while columns (ii) and (iv) represent the nested logit model, which does not incorporate consumer heterogeneity in prices, $\Sigma$.
In Appendix \ref{Appendix_Demand_Estimation}, Table \ref{Demand_Estimates_with_CI} further details the demand estimation for column (i), including 95\% confidence intervals. 
To check the robustness of the nature of competition, the analysis is conducted under two assumptions: “With Hotels,” where Airbnb hosts and hotels are assumed to compete directly \citep{li2019competitive, farronato2022welfare}, and “Only Airbnb,” where hosts and hotels are treated as non-competitors \citep{ForoughifarMehta2024}. 

As shown in Table \ref{Descriptive_Statistics}, all price coefficients are negative ($\alpha < 0$ and $\Sigma < 0$) and significant at the 5\% confidence level, indicating that lower prices significantly enhance consumer utility. The Airbnb dummy is estimated to have a negative effect ($-1.471$ and $-1.391$) on demand, though the standard deviation of this effect is imprecisely estimated. Similarly, the standard errors for ratings are imprecisely estimated. This imprecision is likely due to limited variation in ratings, as nearly 95\% of Airbnb properties maintain average ratings of 4.5 or 5 stars, with properties rated below 3.5 stars being virtually absent \citep{zervas2021first}.

The estimated nest parameters across different assumptions reveal key differences in substitution patterns and the factors influencing consumer decision-making. In the “With Hotels”, the estimated nest parameters ($0.436$ for column (i) and $0.434$ for column (ii)) highlight the significance of both price clusters and the distinction between Airbnb and hotels in consumer decision-making. 
The larger values of the estimated $\rho$ indicate that within-nest substitution plays an important role, underscoring the importance of both price clusters and whether a listing is an Airbnb or a hotel for consumers. Conversely, in the “Only Airbnb”, where nests are determined solely by price clusters, the nest parameters are estimated at $0.132$ and $0.270$ for columns (iii) and (iv), respectively, though their standard errors are imprecise. This suggests that when consumers are restricted to choosing only among Airbnb listings, the substitution pattern within price clusters is weaker, consistent with the smaller absolute value of the price coefficient compared to the “With Hotels” case.

\subsection{Marginal Costs}
I derive marginal costs, new equilibrium prices, and firms' profits under a self-preferencing scenario from the demand estimates and firms' first-order conditions. Table \ref{mc_and_equilibrium_price_descriptive_statistics_with_profits} column (i) presents the marginal cost estimates for RCNL model specifications. 

\begin{figure}[H]
    \centering
    \begin{minipage}{0.49\linewidth}
        \centering
        \includegraphics[width=\linewidth]{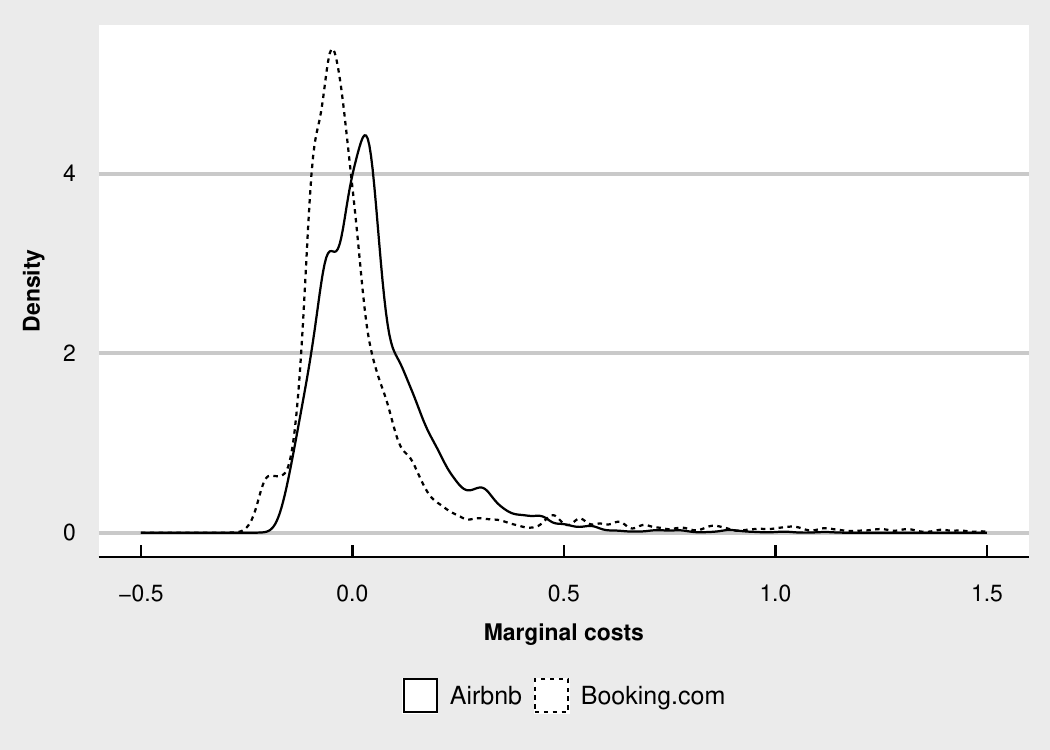}
        \caption*{(a) Marginal Costs by Whether Airbnb}
        \label{fig:ave_price_by_month}
    \end{minipage}
    \hfill
    \begin{minipage}{0.49\linewidth}
        \centering
        \includegraphics[width=\linewidth]{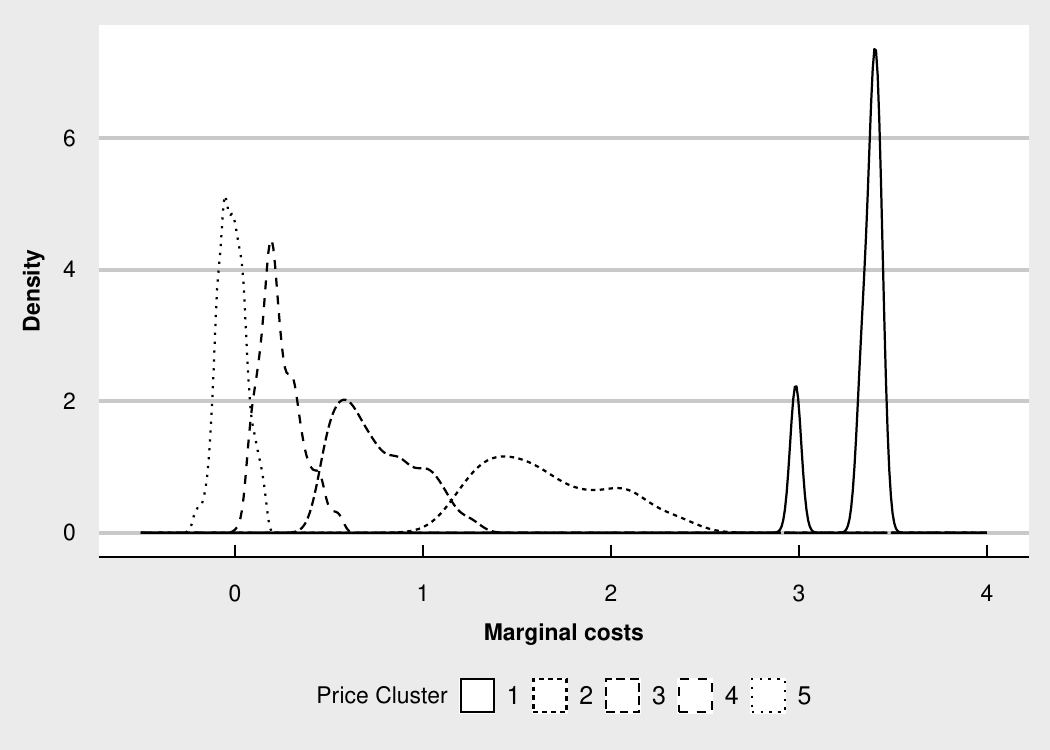}
        \caption*{(b) Marginal Costs by Price Cluster}
        \label{fig:ave_price_by_day_of_week}
    \end{minipage}
    \caption{Marginal Costs, (a) by whether Airbnb and (b) by price cluster.}
     \label{fig:marginal_costs}
    \begin{flushleft}
        \small Note: This figure presents the kernel density estimation results of marginal costs inferred using the RCNL model under the market definition "With Hotels." Panel (a) compares the density estimates between Airbnb and hotels listed on Booking.com, while Panel (b) plots the density estimates by price clusters.
    \end{flushleft}
\end{figure}

The estimated marginal costs reveal intriguing patterns that vary across platform types and price clusters, shedding light on the underlying cost structures despite certain limitations in the model's design. In “With Hotels”, the mean marginal cost for Hosts without SP is $-0.08$, while for hotels, it is $-0.077$, indicating that, on average, hotels have higher marginal costs. 
Similarly, in “Only Airbnb”, the mean marginal cost of hosts without SP is $-0.195$. 
The estimated mean marginal costs are negative, which is counterintuitive. This result arises from my assumption of constant marginal costs, whereby the estimated values are calculated solely based on observed prices and demand.
Figure \ref{fig:marginal_costs} plots the density of estimated marginal costs categorized by (a) whether the listing is on Airbnb and (b) by price cluster. The estimated marginal costs vary significantly across price clusters. 
The result that higher price clusters are associated with higher marginal costs is consistent with \cite{farronato2022welfare}. However, it is important to note that my model defines the market daily and across the 23 wards, which limits the ability to capture supply-side variables that fluctuate daily for each ward. Consequently, the model cannot fully explain the daily variations in marginal costs.

\section{Counterfactual Simulation Results}
\label{counterfactual_simulation_results}
In this section, I discuss the counterfactual simulation result given the estimates of marginal costs and consumer demand. Specifically, I examine the changes in equilibrium prices, profits, and Airbnb's commission revenue, as detailed in Section \ref{supply-side_estimates_counterfactual}, when transitioning from the baseline scenario to the self-preferencing scenario. Following this, in Section \ref{Welfare_Effects}, I evaluate the welfare effects of self-preferencing.

\subsection{Equilibrium Price, Profit, and Airbnb's Commission Revenue}
\label{supply-side_estimates_counterfactual}
In Table \ref{mc_and_equilibrium_price_descriptive_statistics_with_profits}, I present the equilibrium prices and profits for individual hosts and hotels. I also show Airbnb’s total commission revenue for each scenario in Table \ref{Airbnb_commission_rev}.

First, I begin by examining the results under the “With Hotels” assumption. 
When Airbnb engages in self-preferencing, its total commission revenue increases significantly ($+37.726\%$), highlighting its incentive to engage in a self-preferencing strategy. This rise in revenue stems from the differential impact of self-preferencing on hosts and hotels.

For SP hosts, average price increases were modest $(+1.541\%)$, yet they experienced significant profit gains $(+6.542\%)$. 
Non-SP hosts, on the other hand, faced larger average price increases $(+13.053\%)$ but saw only marginal profit improvements $(+0.112\%)$. This discrepancy suggests that the absence of algorithmic support for non-SP hosts leaves them at a strategic disadvantage, limiting their ability to translate higher prices into meaningful profit gains.
Hotels, whose prices remained largely unchanged $(+0.007\%)$, experienced only slight profit improvements $(+0.027\%)$. This stability likely reflects the different competitive dynamics hotels face, as they operate outside Airbnb’s SP framework.
Overall, Airbnb achieved an average revenue increase of $37.726\%$. This significant growth underscores how the platform’s self-preferencing strategy, implemented through the Smart Pricing system, skews the competitive dynamics of the market.

Second, I discuss the results under the “Only Airbnb” assumption and argue the differences from the results under “With Hotels”.
Under “Only Airbnb”, the competitive dynamics shift dramatically, producing outcomes that stand in stark contrast to those observed under “With Hotels”. Notably, Airbnb’s total commission revenue increased by an average of 50.167\%, surpassing the revenue growth achieved in the presence of hotels.

For SP hosts, average price increases were exceptionally high, reaching $+132.417$\%, but these gains did not translate into higher profits. Instead, SP hosts experienced significant profit reductions, averaging $-41.367$\%. This paradox suggests that in a market dominated solely by Airbnb hosts, the algorithm’s adjustments lead to unsustainable price hikes that erode profitability for SP hosts.
Conversely, non-SP hosts adopted a different strategy, reducing prices by 1.862\% on average, which led to modest profit gains of +2.774\%. This behavior underscores the divergent pricing strategies between SP and non-SP hosts under the “Only Airbnb”. Non-SP hosts’ ability to capitalize on relatively lower prices hints at competitive dynamics that differ markedly from those in a mixed market with hotels.
These findings underscore how the presence or absence of hotel competition fundamentally influences market outcomes. While “With Hotels” demonstrates a relatively balanced interplay between SP and non-SP hosts, “Only Airbnb” reveals more pronounced disparities, with Airbnb extracting significantly higher commissions at the expense of host profitability.

\begin{figure}[H]
    \centering
    \includegraphics[width=0.8\linewidth]{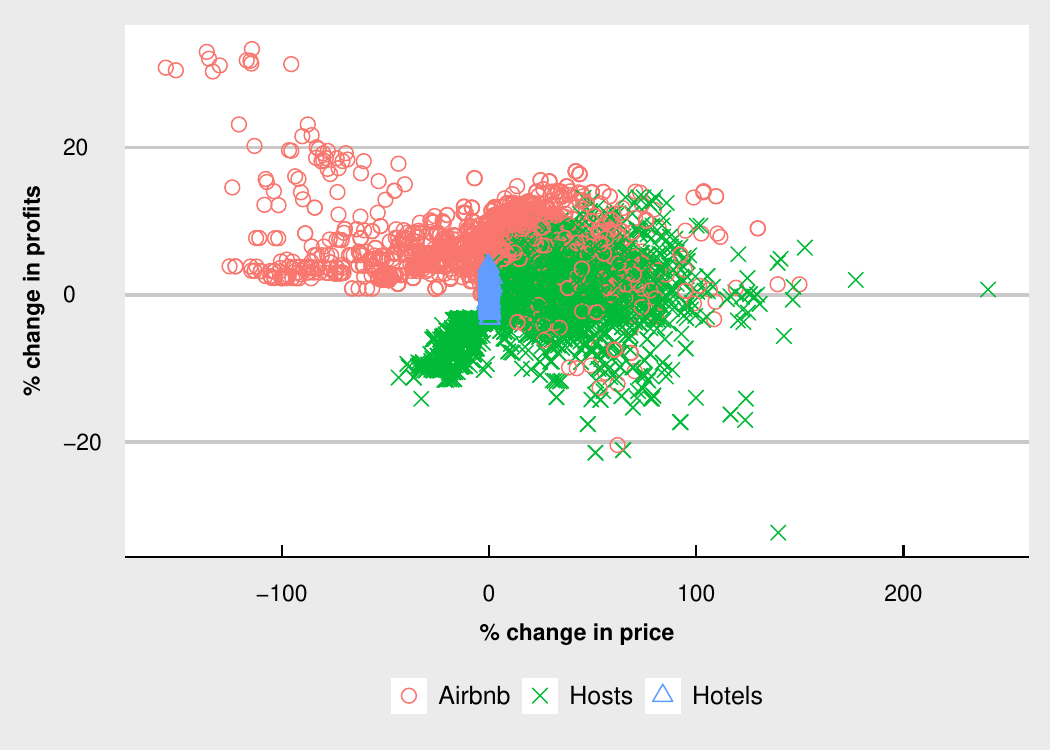}
    \caption{Scatter Plot of \% Change in Price and Profits}
    \label{fig:price_diff_ratio_and_profits}
    \begin{flushleft}
        \small Note: This figure shows the results derived from the RCNL specification for “With Hotels”.
    \end{flushleft}
\end{figure}

To enhance the tabular findings, I present a scatter plot depicting the percentage changes in price and profit under “With Hotels” in Figure \ref{fig:price_diff_ratio_and_profits}.
The analysis reveals that SP hosts (red circles) tend to cluster around moderate price increases or even slight decreases relative to the baseline, yet they achieve substantial profit gains. This pattern highlights Airbnb's strategic pricing adjustments, which optimize aggregate commission revenue through SP. In contrast, non-SP hosts (green crosses) exhibit more varied trends. Some of these hosts have managed to increase profits by slightly lowering prices, a seemingly counterintuitive strategy that effectively counters SP-driven price levels. Meanwhile, hotels (blue triangles) display minimal changes in both prices and profits, with data points concentrated near zero.
These findings align with the tabular results, reinforcing the conclusion that Airbnb’s pricing strategy benefits SP hosts the most while preserving stable outcomes for hotels.

\begin{figure}[H]
    \centering
    \includegraphics[width=0.7\linewidth]{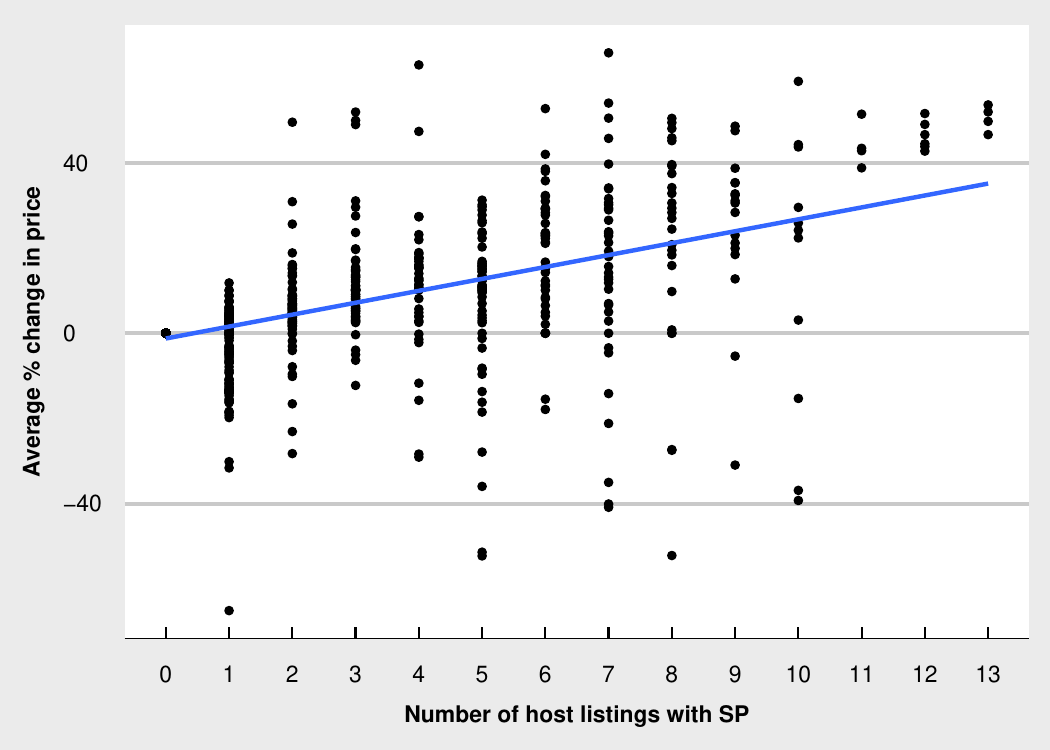}
    \caption{Scatter Plot of Average \% Change in Price and the Number of Host listings with SP}
    \label{fig:average_price_diff_and_number_of_host_with_SP}
    \begin{flushleft}
        \small Note: This figure shows the average rate of change of the equilibrium price in the market, defined as “With Hotels,” when moving from the baseline scenario to the self-preferencing scenario on the y-axis and the number of hosts using SP per market on the x-axis. The average price change is calculated for the combined sample of hosts on Airbnb and hotels on Booking.com.
    \end{flushleft}
\end{figure}

\begin{figure}[H]
    \centering
    \begin{minipage}{0.49\linewidth}
        \centering
        \includegraphics[width=\linewidth]{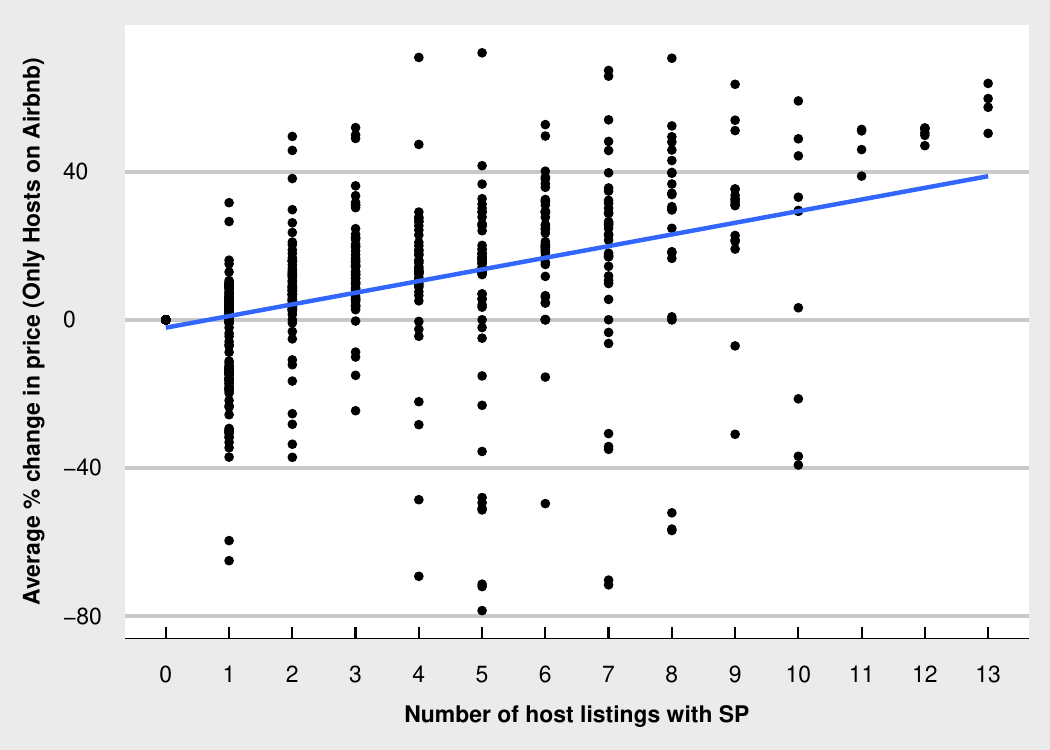}
        \caption*{(a) Only Listings on Airbnb}
    \end{minipage}
    \hfill
    \begin{minipage}{0.49\linewidth}
        \centering
        \includegraphics[width=\linewidth]{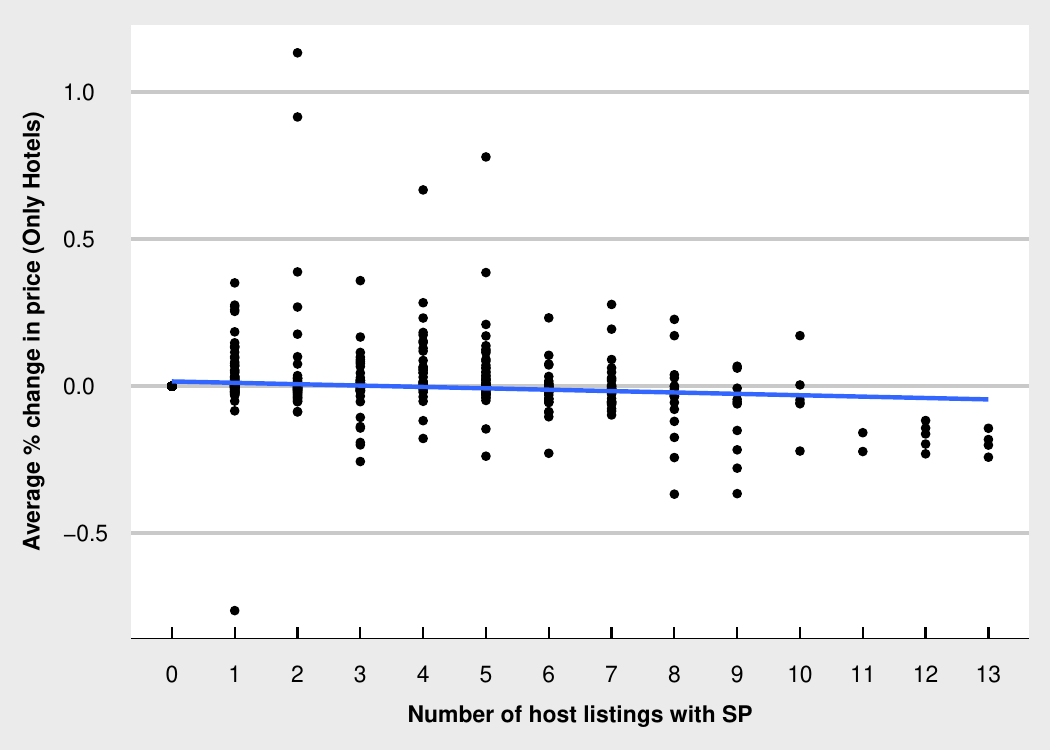}
        \caption*{(b) Only Hotels on Booking.com}
    \end{minipage}
    \caption{Scatter Plot of Average \% Change in Price and the Number of Host listings with SP}
     \label{fig:ave_price_change_separating_by_whether_Airbnb}
    \begin{flushleft}
        \small Note: This figure represents the percentage change in price across scenarios within the market defined as “With Hotels,” plotted separately for (a) only host listings on Airbnb and (b) only hotels.
    \end{flushleft}
\end{figure}

In addition, I found that Airbnb's self-preferencing mitigates price competition as the number of SP hosts in the market increases. To capture this effect, I provide scatter plots in Figure \ref{fig:average_price_diff_and_number_of_host_with_SP} and Figure \ref{fig:ave_price_change_separating_by_whether_Airbnb}, illustrating the relationship between the number of SP hosts per market and the average rate of price change across different scenarios under “With Hotels”.
As the number of hosts with SP increases, hosts' prices rise, and competition among hosts diminishes, while competition between Airbnb and hotels remains relatively stable. This intuitive result highlights significant heterogeneity in the impact of self-preferencing, emphasizing its role in reducing price competition.

In summary, Airbnb’s shift to platform-centric pricing has led to divergent outcomes across SP hosts, non-SP hosts, and hotels. These findings demonstrate Airbnb’s strong short-run incentives to engage in self-preferencing, as it allows the platform to efficiently maximize commission revenues and capture a larger share of the market surplus.

\subsection{Welfare Effects of Self-Preferencing}
\label{Welfare_Effects}
This subsection answers the research question of this paper, ``What is the impact of self-preferencing by the platform on social welfare?” using the estimated models. In Table \ref{Welfare_effects}, I compare consumer surplus (CS), producer surplus (PS), and social welfare (SW) between two assumptions.

Under the “With Hotels” assumption, consumers bear substantial losses as consumer surplus declines by an average of 5.077\% when Airbnb shifts from a baseline scenario to a self-preferencing scenario. While producer surplus increases by an average of 12.383\%, the decline in consumer surplus translates into a noticeable reduction in social welfare, which drops by 5.077\% relative to the baseline scenario. This outcome highlights a fundamental imbalance: while Airbnb may achieve huge gains, they are far outweighed by the significant losses in consumer welfare.

The contrast is even starker in the “Only Airbnb” assumption. On average, consumer surplus declines by 11.933\%, effectively negating the 30.962\% rise in producer surplus. The resulting net effect is a pronounced reduction in social welfare. While Airbnb does succeed in extracting higher commission revenues, this comes at a clear cost to consumers and the market's overall efficiency.

How significant is this welfare loss for the entire vacation rental market in Tokyo? Although my data spans only four months from March to July 2024, a projection of the estimated daily social welfare loss in Tokyo's 23 wards (2.65 million yen) over a full year reveals the gravity of the situation. Multiplying the daily loss by 365 yields an annual social welfare loss of 968.345 million yen due to Airbnb's self-preferencing.
To put this into perspective, the vacation rental market in Tokyo was valued at approximately 6,500 million yen in 2023.\footnote{In 2023, the estimated total consumer spending in the vacation rental market across Japan was approximately 26 billion yen (based on the Japan Tourism Agency's Survey of Travel and Tourism Consumption Trends, \url{https://www.mlit.go.jp/kankocho/tokei_hakusyo/shohidoko.html}). During the same year, the total number of users of budget accommodations in Tokyo was approximately 22.88 million accounting for 25\% of the national total, according to the Japan Tourism Agency's Statistical Survey on Overnight Travel, \url{https://www.mlit.go.jp/kankocho/tokei_hakusyo/shukuhakutokei.html}. Based on this proportion, the size of Tokyo's vacation rental market in 2023 is calculated as 26 billion yen multiplied by 0.25, resulting in approximately 6.5 billion yen.} Thus, this loss constitutes about 14.90\% of the market’s total social welfare—a substantial impact. This considerable welfare loss emphasizes the urgent need for regulatory intervention through competition policy to mitigate the adverse effects of self-preferencing.

The main takeaway is that while self-preferencing can enhance the platform’s revenue, they do so at the expense of consumer welfare and overall economic efficiency. From a policy perspective, this conclusion suggests that competition authorities need to pay close attention to the opaque nature of algorithmic decision-making and pricing practices. Since these algorithms are proprietary and remain hidden from public scrutiny, policymakers may need to consider targeted interventions to improve transparency and accountability. This could involve mandating at least partial disclosures of algorithmic objectives or outcomes and introducing guidelines that limit certain forms of strategic self-preferencing. Such policies would aim to ensure that platforms cannot unilaterally reshape the competitive landscape in ways that harm consumers and reduce social welfare.
\section{Conclusion}
\label{Conclusion}
I analyzed the welfare effects of self-preferencing by Airbnb in this paper. Specifically, when Airbnb implements self-preferencing by offering hosts a pricing algorithm (SP) designed to maximize the platform's total commission revenue, my analysis reveals a significant negative impact on social welfare, with an average decline of 5.08\%. At the same time, Airbnb's total commission revenue increases by 37.73\% on average, highlighting a strong incentive for the platform to engage in self-preferencing. These findings have significant policy implications, particularly in light of the widespread lack of transparency and accountability regarding the algorithm they employ. Although I focused on Airbnb, my findings can be applied more generally to marketplace platforms such as Amazon and Mercari which provide the pricing algorithm to sellers. 

This study has several limitations. First, while I analyze the impact of self-preferencing on welfare through the incentives of platforms, I do not capture the potential positive effects of pricing algorithms. My model especially does not account for demand forecasting improvements and pricing cost reductions \citep{miklos2019collusion, Connor-Wilson-2021reduced}. 

Second, this study examines the short-run incentives for the platform to engage in self-preferencing, but it highlights potential long-run trade-offs that warrant further investigation. While my findings suggest that self-preferencing can benefit the platform in the short run, it also reduces the profitability of some platform users. In the long run, the platform's profitability is likely related to increasing the number of users on its platform \citep{belleflamme2021economics}. If self-preferencing leads to declining host profitability, it may ultimately discourage participation and reduce the user base, creating a significant trade-off for the platform. These interactions imply that the platform’s behavior could be better understood through long-run dynamics. Developing a model incorporating these dynamics could provide valuable insights into how self-preferencing impacts the platform’s growth strategy and relationship with hosts.

Finally, I do not capture the strategic behavior of consumers in response to dynamic pricing  \citep{hendel2013intertemporal, noguchi2024_jmp}, which could result in different welfare effects due to Airbnb's self-preferencing through SP. A promising avenue for future research is to investigate dynamic pricing games where consumers strategically either expedite or delay their purchasing decisions, potentially influencing pricing outcomes and welfare implications of self-preferencing.

\addcontentsline{toc}{section}{References}
\bibliographystyle{econ-aea}
\bibliography{ref}

\newpage
\appendix
\section{Additional explanations and analysis}
\subsection{Detailed Explanation of Data}
\label{data_explanation}
In this appendix, I explain the reason why I selected Booking.com, the methodology used to calculate vacancies and the number of beds in Airbnb and hotel listings, and the adjustments made to ensure comparability between the two data sets.

Why did I obtain hotel information from Booking.com? As of December 27, 2023, I calculated that approximately 17\% of the reviews for Airbnb listings in Tokyo were written in Japanese, based on data I retrieved from Inside Airbnb.\footnote{\url{https://insideairbnb.com/}, accessed on 29 April 2024.}
This indicates that the majority of Airbnb users in Tokyo are foreign guests. Therefore, hotel information was collected from Booking.com, the most widely used hotel booking platform worldwide.\footnote{A information from a website, \textit{similarweb}: \url{https://www.similarweb.com/top-websites/travel-and-tourism/accommodation-and-hotels/}, accessed on 7 December 2024.}

The vacancies in Airbnb listings were calculated as follows: Listings supplied by the same host at the same price within the same market were treated as identical properties. This approach accounts for cases where hosts own multiple listings differentiated only by room numbers. Such listings were considered as a single property, and vacancies were counted cumulatively to compute market shares. For hotels on Booking.com, the data exclusively includes rooms without breakfast, considering that Airbnb accommodations typically do not provide breakfast. 

I also quantified the number of beds. In accommodations such as vacation rentals and hotels, various beds are available, including Single, Semi-Double, Double, Queen, King, Bunk, Sofa, and Futon. Table \ref{how_to_count_num_of_bed} presents detailed descriptions, sizes, and the relative width of each bed type, standardized by setting the width of a single bed as the baseline. To treat the number of beds as a continuous variable, I calculated the "Relative Value to Single Bed" which is calculated based on the bed widths using the formula "width / 90" where 90 is the width of a single bed.

\subsection{Demand Model and Welfare Measure}
\label{appendix_demand_model_and_welfare_measure}
This Appendix section supplements section \ref{Consumer_Demand_Model} (Consumer Demand). I indicate the computation of the choice probability (i.e., the demand) for product $j$ in market $m$ and the calculation of consumer welfare in the RCNL model.

There are $G+1$ nesting groups ($g \in \{0, 1, 2, \ldots, G\} := \mathcal{G}$), and each product $j$ is assigned to one of these groups. The outside good is assigned to group $0$. The set $\mathcal{J}_{gm} \subset \mathcal{J}_m$ denotes the products in group $g$ and market $m$.
The indirect utility function \eqref{indirect-utility-specification} is rewritten as follows:  
\begin{align*}
    u_{ijm} &= (\alpha + \Sigma {v}_{im}) p_{jm} + \mathbf{X}'_{jm}\boldsymbol{\beta} + \xi_{jm} + \bar{\varepsilon}_{ijm} \\
    &= \delta_{jm} + \mu_{ijm} + \zeta_{igm} + (1-\rho)\varepsilon_{ijm} 
\end{align*}
where
\begin{align*}
    \delta_{jm} &= \alpha p_{jm} + \mathbf{X}'_{jm}\boldsymbol{\beta} + \xi_{jm}\\
    \mu_{ijm} &= \Sigma v_{im}p_{jm}.
\end{align*}
Then, the market share of product $j$ in market $m$ is computed as
\begin{equation*}
    s_{jm} = \int\frac{\exp[\delta_{jm} + \mu_{ijm}/(1-\rho)]}{\exp[IV_{igm}/(1-\rho)]}\cdot\frac{\exp IV_{igm}}{1 + \sum_{g'\in \mathcal{G}}\exp IV_{igm}}f(\mu_{im}\mid\Sigma^2)d\mu_{im}.
\end{equation*}
where $\mu_{im} = \qty(\mu_{ijm})_{j\in\mathcal{J}_m}$, $f(\cdot)$ is the probability density function of $\mu_{im}$, and
\begin{equation*}
    IV_{igm} = (1-\rho)\log\sum_{j\in\mathcal{J}_{gm}}\exp\qty(\frac{\delta_{jm} + \mu_{ijm}}{1-\rho})
\end{equation*}
is the inclusive value term, which reflects the expected utility derived from the alternatives within group $g$. Based on the above, the demand for product $j$ in market $m$, $d_{jm}$, can be obtained by multiplying the market share by the market size $I_m$, i.e., $d_{jm} = s_{jm} \cdot I_m$.

Finally, the consumer surplus for individual $i$ is shown to be calculated, following \cite{mcfadden1981econometric} and \cite{small1981applied}, as
\begin{equation*}
    \text{CS}_{im} = \log\qty(1 + \sum_{g\in \mathcal{G}} \exp IV_{igm}) / (\alpha + \Sigma v_{im}).
\end{equation*}
Producer surplus can be obtained by summing up the profits of firms in each market. The sum of consumer surplus and producer surplus represents social welfare.

\subsection{Demand Estimation}
\label{Appendix_Demand_Estimation}
In this section, I provide the estimated results of the main interest model specification with 95\% confidence intervals. The confidence intervals are computed using a parametric bootstrap procedure with 1000 resampled. The parametric bootstrap procedure is as follows: (i) construct 1000 draws from the estimated joint distribution of parameters assuming the joint distribution of parameters follows the normal distribution, (ii) compute the implied mean utility, shares, marginal costs, and prices for each draw, and (iii) calculate the required post-estimation output for all parametric bootstrap samples. Using the resulting empirical distribution, construct bootstrap confidence intervals. I use \textit{PyBLP} to construct these confidence intervals, see \cite{PyBLP} for details.

\begin{table}[h]
    \centering
    \caption{Demand Estimates of RCNL model with Confidence Intervals}
    \label{Demand_Estimates_with_CI}
    \begin{tabular}{lccc}
    \toprule
    Parameter & Coeff. & Std. err. & 95\% C.I. \\
    \midrule
    Price ($\alpha$)            & $-0.928$ & $0.334$ & $[-1.61, -0.298]$ \\
    Price ($\Sigma$)            & $-0.559$ & $0.077$ & $[-0.706, -0.405]$  \\
    Size nests ($\rho$)         & $0.436$  & $0.154$ & $[0.134, 0.734]$   \\
    \# Beds                     & $-0.005$  & $0.077$ & $[-3.86, 0.839]$    \\
    \# Reviews                  & $-0.00035$ & $0.00040$ & $[-0.00120, 0.00042]$ \\
    Airbnb Dummy                & $-1.471$ & $1.180$ & $[-0.159, 0.144]$ \\
    Ratings of Room             & $-0.486$ & $1.151$ & $[-0.459, 1.192]$  \\
    Ratings of Cleanness        & $0.358$  & $0.425$ & $[-0.83, 3.00]$   \\
    Ratings of Location         & $-0.390$  & $0.569$ & $[-1.28, 2.46]$  \\
    Ratings of Staff Communication & $0.608$ & $0.940$ & $[-2.74, 1.84]$  \\
    City Fixed Effect           & \multicolumn{3}{c}{Yes} \\
    Month Fixed Effect          & \multicolumn{3}{c}{Yes} \\
    Day of the Week Fixed Effect & \multicolumn{3}{c}{Yes} \\
    Robust S.E. for Cluster     & \multicolumn{3}{c}{Yes} \\
    \# Markets                  & \multicolumn{3}{c}{$899$} \\
    N                           & \multicolumn{3}{c}{$17607$} \\
    \bottomrule
    \end{tabular}
    \begin{flushleft}
    \small Note: This table shows the results of demand estimation by data containing all clusters from 1 to 5. I constructed the confidence interval by parametric bootstrapping using \textit{PyBLP}. Standard errors are clustered by price cluster.
    \end{flushleft}
\end{table}

\section{Figures}

\section{Tables}

\begin{sidewaystable}
\caption{Descriptive Statistics}
\label{Descriptive_Statistics}
    \centering
    \begin{tblr}[         
        ]                     
        {                     
        colspec={Q[]Q[]Q[]Q[]Q[]Q[]Q[]},
        cell{1}{2}={c=2,}{halign=c,},
        cell{1}{4}={c=2,}{halign=c,},
        column{1}={halign=l,},
        column{2}={halign=r,},
        column{3}={halign=r,},
        column{4}={halign=r,},
        column{5}={halign=r,},
        column{6}={halign=r,},
        column{7}={halign=r,},
        row{1}={halign=c,},
        }                     
        \toprule
        & Airbnb (N=54988) &  & Booking.com (N=86388) &  &  &  \\ \cmidrule[lr]{2-3}\cmidrule[lr]{4-5}
        & Mean & SD & Mean & SD & Diff. in Means & SE \\ \midrule 
        Prices                        & \num{0.287}  & \num{0.191}  & \num{0.337}    & \num{0.420}    & \num{0.050}***    & \num{0.002} \\
        \# Reviews                   & \num{47.755} & \num{69.823} & \num{1450.447} & \num{1584.243} & \num{1402.692}*** & \num{5.398} \\
        Rating of Location            & \num{4.635}  & \num{0.265}  & \num{4.113}    & \num{0.386}    & \num{-0.522}***   & \num{0.002} \\
        Rating of Staff Communication & \num{4.769}  & \num{0.219}  & \num{4.265}    & \num{0.368}    & \num{-0.504}***   & \num{0.002} \\
        Rating of Cleanness           & \num{4.603}  & \num{0.327}  & \num{4.150}    & \num{0.455}    & \num{-0.453}***   & \num{0.002} \\
        Rating of Room                & \num{4.599}  & \num{0.288}  & \num{4.140}    & \num{0.385}    & \num{-0.458}***   & \num{0.002} \\
        \# Vacancies                 & \num{4.241}  & \num{4.522}  & \num{4.749}    & \num{3.530}    & \num{0.509}***    & \num{0.023} \\
        \# Beds                      & \num{2.155}  & \num{1.380}  & \num{1.936}    & \num{1.242}    & \num{-0.219}***   & \num{0.007} \\
        \bottomrule
\end{tblr}
    \begin{flushleft}
    \small Note: Prices are scaled by dividing them by 100,000. Additionally, while the ratings on Booking.com were originally rated on a 10-point scale, they have been rescaled to a 1-5 scale to align with Airbnb's rating system. \\$*p<0.1; **p<0.05; ***p<0.01$.
    \end{flushleft}
\end{sidewaystable}

\begin{sidewaystable}
\centering
\caption{Main Types of Beds and Their Characteristics}
\label{how_to_count_num_of_bed}
\begin{tabular}{|l|p{7cm}|c|c|}
\hline
\textbf{Type} & \textbf{Description} & \textbf{Size (Width × Length cm)} & \textbf{Relative Value to Single Bed} \\ \hline
Single Bed & The smallest size bed for one person. & 90 × 190 & 1 \\ \hline
Semi-Double Bed & Wider than a single bed, offering more comfort for one person. & 120 × 190 & 1.33 \\ \hline
Double Bed & A standard-size bed suitable for two people. & 140 × 190 & 1.56 \\ \hline
Queen Bed & Larger than a double bed, providing more comfort for two people. & 160 × 200 & 1.78 \\ \hline
King Bed & A spacious bed close to the largest size. & 180 × 200 & 2 \\ \hline
Bunk Bed & Two single beds stacked vertically. & Each 90 × 190 & 2 \\ \hline
Sofa Bed & A versatile bed that doubles as a sofa and converts into a bed. & 90 × 190 & 1 \\ \hline
Futon & A traditional Japanese bedding that can be folded and stored. & None & 1 \\ \hline
\end{tabular}
\label{tab:main_bed_types}
\begin{flushleft}
    \small Note: This table presents the types of beds available in accommodations listed on Airbnb and hotels, along with their textual descriptions, sizes, and widths normalized relative to a single bed, where the width of a single bed is set to 1.
\end{flushleft}
\end{sidewaystable}

\setstretch{1.2}
\begin{table}[H]
    \caption{Results of K-Means Clustering on Prices}
    \label{K-means}
    \centering
    \begin{tabular}[t]{lrllllll}
        \toprule
        Cluster & N & Airbnb (\%) & Hotels (\%) & Mean & SD & Min & Max\\
        \midrule
        1 & 889 & 1.912 & 98.088 & 4.096 & 1.107 & 2.975 & 7.900\\
        2 & 5391 & 0.556 & 99.444 & 1.888 & 0.371 & 1.440 & 2.965\\
        3 & 12119 & 14.250 & 85.750 & 1.012 & 0.202 & 0.735 & 1.439\\
        4 & 46513 & 30.931 & 69.069 & 0.460 & 0.112 & 0.314 & 0.734\\
        5 & 133995 & 29.176 & 70.824 & 0.170 & 0.069 & 0.016 & 0.314\\
        \bottomrule
    \end{tabular}
    \begin{flushleft}
        \small Note: I set the total number of clusters to 5 following past studies \citep{zervas2017rise, farronato2022welfare}. Cluster 1 represents the highest price range, while Cluster 5 represents the lowest price range.
    \end{flushleft}
\end{table}
\doublespacing
\noindent

\setstretch{1.2}
\begin{table}[h]
    \caption{SP ratio}
    \label{SP_ratio}
    \centering
    \begin{tabular}[t]{rrrll}
    \toprule
    Cluster & \# SP & \# Non-SP & SP (\%) & Non-SP (\%)\\
    \midrule
    1 & 0 & 17 & 0.000 & 100.000\\
    2 & 0 & 30 & 0.000 & 100.000\\
    3 & 0 & 1724 & 0.000 & 100.000\\
    4 & 569 & 13767 & 3.969 & 96.031\\
    5 & 8903 & 29978 & 22.898 & 77.102\\
    \bottomrule
    \end{tabular}
    \begin{flushleft}
        \small Note: This table shows the number and proportion of hosts who meet both criteria for utilizing Smart Pricing: (1) Operating Low-Priced Listings and (2) Engaging in Dynamic Pricing, categorized by cluster. Data from hotels listed on Booking.com are excluded from the calculations.
    \end{flushleft}
    \end{table}
    
\begin{sidewaystable}
\caption{Descriptive statistics of datasets used to estimate the structural model}
\label{Descriptive_statistics_for_estimation}
\centering
\begin{tblr}[         
]                     
{                     
colspec={Q[]Q[]Q[]Q[]Q[]Q[]Q[]},
cell{1}{2}={c=2,}{halign=c,},
cell{1}{4}={c=2,}{halign=c,},
column{1}={halign=l,},
column{2}={halign=r,},
column{3}={halign=r,},
column{4}={halign=r,},
column{5}={halign=r,},
column{6}={halign=r,},
column{7}={halign=r,},
row{1}={halign=c,},
}                     
\toprule
& Airbnb (N=9194) &  & Booking.com (N=8413) &  &  &  \\ \cmidrule[lr]{2-3}\cmidrule[lr]{4-5}
& Mean & SD & Mean & SD & Diff. in Means & Std. Error \\ \midrule 
Prices                        & \num{0.245}  & \num{0.142}  & \num{0.218}    & \num{0.324}    & \num{-0.027}***   & \num{0.004}  \\
\# Reviews                   & \num{48.519} & \num{72.608} & \num{1098.093} & \num{1179.268} & \num{1049.574}*** & \num{12.879} \\
Rating of Location            & \num{4.598}  & \num{0.290}  & \num{3.913}    & \num{0.390}    & \num{-0.684}***   & \num{0.005}  \\
Rating of Staff Communication & \num{4.762}  & \num{0.213}  & \num{4.148}    & \num{0.402}    & \num{-0.614}***   & \num{0.005}  \\
Rating of Cleanness           & \num{4.602}  & \num{0.304}  & \num{3.977}    & \num{0.507}    & \num{-0.626}***   & \num{0.006}  \\
Rating of Room                & \num{4.608}  & \num{0.247}  & \num{3.985}    & \num{0.432}    & \num{-0.623}***   & \num{0.005}  \\
\# Vacancies                 & \num{6.000}  & \num{5.391}  & \num{7.537}    & \num{3.017}    & \num{1.537}***    & \num{0.065}  \\
\# Beds                      & \num{2.144}  & \num{1.144}  & \num{1.621}    & \num{0.933}    & \num{-0.523}***   & \num{0.016}  \\
\bottomrule
\end{tblr}
\begin{flushleft}
    \small Note: The descriptive statistics presented in this table pertain to the dataset used for model estimation. Specifically, only data with calculated market shares of 0.5\% or higher are included.
\end{flushleft}
\end{sidewaystable}

\begin{table}[H]
    \centering
    \caption{The Number of Firms in Each Market}
    \label{The_Number_of_Firms_in_Each_Market}
    \begin{tblr}[         
        ]                     
        {                     
        colspec={Q[]Q[]Q[]Q[]Q[]Q[]Q[]},
        column{1}={halign=l,},
        column{2}={halign=r,},
        column{3}={halign=r,},
        column{4}={halign=r,},
        column{5}={halign=r,},
        column{6}={halign=r,},
        column{7}={halign=r,},
        }                     
        \toprule
        & Mean & SD & Min & Max & Median & N \\ \midrule 
        Hosts Without SP & \num{3.228} & \num{3.546} & \num{0.000} & \num{25.000} & \num{2.000} & 899 \\
        Hosts With SP       & \num{2.304} & 2.961 & \num{0.000} & 13.000  & \num{1.000} & 899 \\
        Hotels       & \num{4.050} & \num{4.015} & \num{0.000} & \num{21.000} & \num{3.000} & 899 \\
        \midrule
        Total        & \num{7.278} & \num{4.068} & \num{1.000} & \num{25.000} & \num{7.000} & 899 \\
        \bottomrule
    \end{tblr}
    \begin{flushleft}
        \small Note: This table shows descriptive statistics of the number of hosts and hotels per market (pairs of one day and one of Tokyo's 23 wards) and their totals after applying the sample restrictions for demand estimation. N represents the number of markets, not the number of listings.
    \end{flushleft}
\end{table}

\setstretch{1.2}
\begin{table}[H]
    \caption{Strategic Environment for Self-preferencing Scenario}
    \label{scenario2-game-environment}
    \centering
    \begin{talltblr}[
        entry=none,label=none,
        ]
        {
        colspec={Q[]Q[]Q[]},
        cell{1}{1}={halign=l,},
        cell{1}{2}={halign=l,},
        cell{1}{3}={halign=l,},
        column{1}={halign=l,},
        column{2}={halign=c,},
        column{3}={halign=c,},
        row{1}={halign=c,},
        }
        \toprule
        Player ($f$) & Set of goods & Max. Problem \\
        \midrule
        Airbnb & $\mathcal{J}_{\text{SP}}$ & $\max\limits_{p_j}\tau\qty(\sum\limits_{j~\in ~\mathcal{J}_{\text{Airbnb}}}~p_jq_j)~~\forall j\in\mathcal{J}_{\text{SP}}$ \\
        Non-SP hosts & $\mathcal{J}_f \subset \mathcal{J}_{\text{non-SP}}$ & $\max\limits_{p_j}~\sum\limits_{j\in \mathcal{J}_f}(p_{j}-\text{mc}_{j})q_{j}~~\forall j \in \mathcal{J}_f$ \\
        Hotels & $\mathcal{J}_f \subset \mathcal{J}_{\text{hotel}}$ & Same as for Non-SP hosts \\ 
        \bottomrule
    \end{talltblr}
    \begin{flushleft}
        \small
        Notes: $\tau$ denotes the commission charged by the platform firm to platform users (i.e., $\tau = 0.03$). $\mathcal{J}_{\text{SP}}$ refers to the set of product prices for SP-adopting hosts on Airbnb. $\mathcal{J}_{\text{Airbnb}}$ represents the set of all hosts' products on Airbnb. {Adoption of SP} is determined based on whether the product belongs to the ”low-price range (clusters 4 or 5).”
    \end{flushleft}
\end{table}
\doublespacing

\setstretch{1}
\begin{table}[H]
    \centering
    \caption{Demand Estimates}
    \label{Demand_Estimates}
    \begingroup
    \begin{tabularx}{\textwidth}{@{}l *{4}{X}@{}}
    \toprule
    & \multicolumn{2}{c}{With Hotels} & \multicolumn{2}{c}{Airbnb Only} \\ 
    \cmidrule(lr){2-3} \cmidrule(lr){4-5}
    Parameters & (i) RCNL & (ii) NL & (iii) RCNL & (iv) NL \\ 
    \midrule
    Price ($\alpha$)              & $-0.928$ & $-0.852$ & $-0.468$ & $-0.144$ \\
                                   & ($0.334$) & ($0.303$) & ($0.027$) & ($0.046$) \\ \addlinespace
    Price ($\Sigma$)              & $-0.559$ & $-$ & $-2.809$ & $-$ \\
                                   & ($0.077$) &  & ($0.133$) &  \\ \addlinespace
    Size nests ($\rho$)           & $0.436$ & $0.434$ & $0.132$ & $0.270$ \\
                                   & ($0.154$) & ($0.146$) & ($0.464$) & ($0.598$) \\ \addlinespace
    \# Beds                       & $-0.005$ & $-0.030$ & $-0.173$ & $-0.206$ \\
                                   & ($0.077$) & ($0.063$) & ($0.101$) & ($0.227$) \\ \addlinespace
    \# Reviews                    & $-0.00035$ & $-0.00033$ & $0.00025$ & $0.00055$ \\
                                   & ($0.00040$) & ($0.00038$) & ($0.00016$) & ($0.00139$) \\ \addlinespace
    Airbnb Dummy                  & $-1.471$ & $-1.391$ & $-$ & $-$ \\
                                   & ($1.180$) & ($1.133$) &  &  \\ \addlinespace
    Ratings of Room               & $-0.486$ & $-0.394$ & $-0.454$ & $-1.965$ \\
                                   & ($1.151$) & ($1.126$) & ($0.131$) & ($0.296$) \\ \addlinespace
    Ratings of Cleanness          & $0.358$ & $0.282$ & $-0.378$ & $-1.906$ \\
                                   & ($0.425$) & ($0.413$) & ($0.162$) & ($0.268$) \\ \addlinespace
    Ratings of Location           & $-0.390$ & $-0.310$ & $-0.364$ & $0.041$ \\
                                   & ($0.569$) & ($0.543$) & ($0.183$) & ($0.306$) \\ \addlinespace
    Ratings of Staff Communication & $0.608$ & $0.533$ & $1.302$ & $-0.368$ \\
                                   & ($0.940$) & ($0.902$) & ($0.386$) & ($0.106$) \\ \addlinespace
    City Fixed Effect           & Yes & Yes & Yes & Yes \\
    Month Fixed Effect          & Yes & Yes & Yes & Yes \\
    Day of the Week Fixed Effect & Yes & Yes & Yes & Yes \\
    \# Markets & $899$ & $899$ & $1436$ & $1436$ \\
    N                 & $17607$ & $17607$ & $23667$ & $23667$ \\
    \midrule[\heavyrulewidth]
    \end{tabularx} \endgroup
    \begin{flushleft}
        \small{Note: This table shows the results of demand estimation. RCNL and NL represent random coefficient nested logit, and nested logit, respectively. Robust for a nesting group standard errors are in parentheses.}
    \end{flushleft}
\end{table}
\doublespacing

\setstretch{1.2}
\begin{table}[H]
    \centering
    \caption{Descriptive Statistics of Supply-Side Estimates (RCNL)}
    \label{mc_and_equilibrium_price_descriptive_statistics_with_profits}
    \begin{tblr}{
        colspec={Q[]Q[]Q[]Q[]Q[]},
        column{1}={halign=l},
        column{2}={halign=l},
        column{3}={halign=r,},
        column{4}={halign=r,},
        column{5}={halign=r,},
    }
        \toprule
         &   & (i) MC & (ii) \% change in price & (iii) \% change in profits \\ \midrule
         A. With Hotels \\
        Hosts with SP & Mean & $-$  & \num{1.541}   & \num{6.542}   \\
        & SD   & $-$  & \num{40.453}  & \num{4.718}   \\
        & Min  & $-$ & \num{-155.933} & \num{-20.447} \\
        & Max  & $-$  & \num{149.767} & \num{33.384}  \\
        \midrule
        Hosts without SP & Mean & \num{-0.080}  & \num{13.053}   & \num{0.112}   \\
        & SD   & \num{0.162}  & \num{24.038}  & \num{3.560}   \\
        & Min  & \num{-0.406} & \num{-43.590} & \num{-32.359} \\
        & Max  & \num{0.968}  & \num{240.815} & \num{13.248}  \\
        \midrule
        Hotels & Mean & \num{-0.077}  & \num{0.007}   & \num{0.027}   \\
         & SD   & \num{0.292}  & \num{0.064}   & \num{0.482}   \\
        & Min  & \num{-0.479} & \num{-1.077}  & \num{-3.202}  \\
        & Max  & \num{3.081}  & \num{1.734}   & \num{3.950}   \\
        \midrule
        B. Only Airbnb \\
        Hosts with SP & Mean & $-$ & \num{132.417} & \num{-41.367} \\
         & SD   & $-$  & \num{61.002}  & \num{8.676}   \\
        & Min  & $-$ & \num{15.618}  & \num{-63.924} \\
        & Max  & $-$ & \num{463.926} & \num{-2.077}  \\
        \midrule
        Hosts without SP & Mean & \num{-0.195} & \num{-1.862}  & \num{2.774}  \\
        & SD   & \num{0.150}  & \num{8.587}   & \num{3.619}   \\
        & Min  & \num{-0.442} & \num{-78.074} & \num{-0.146}  \\
        & Max  & \num{1.377}  & \num{18.888}  & \num{28.373}  \\
        \bottomrule
    \end{tblr}
    \begin{flushleft}
        \small Note: MC stands for marginal costs. This table shows the marginal cost, percentage change in the equilibrium price, and percentage change in profits under the self-preferencing scenario from the price under the baseline scenario using RCNL specifications. The marginal cost of hosts using SP is assumed to be zero, so their marginal costs are not estimated. The percentage changes in prices and profits represent the extent to which prices and profits have changed from the Baseline scenario to the Self-Preferencing scenario.
    \end{flushleft}
\end{table}

\doublespacing

\begin{table}[H]
    \centering
    \caption{Airbnb's Commission Revenue per Market}
    \label{Airbnb_commission_rev}
    \begin{tblr}[         
    ]                     
    {                     
    colspec={Q[]Q[]Q[]Q[]Q[]},
    column{1}={halign=l,},
    column{2}={halign=r,},
    column{3}={halign=r,},
    column{4}={halign=r,},
    column{5}={halign=r,},
    }                     
    \toprule
    & Mean & SD & Min & Max \\ \midrule 
    A. With Hotels\\
    Commission Revenue (baseline)                & \num{0.501}  & \num{0.848}   & \num{0.000}   & \num{9.878}    \\
    Commission Revenue (self-preferencing)          & \num{0.646}  & \num{1.210}   & \num{0.000}   & \num{12.524}   \\
    \% change in Commission Revenue & \num{37.726} & \num{184.645} & \num{-98.135} & \num{3000.000} \\
    \midrule
    B. Only Airbnb\\
    Commission Revenue (baseline)                & \num{0.517}  & \num{1.180}   & \num{0.006}   & \num{7.841}    \\
    Commission Revenue (self-preferencing)          & \num{0.856}  & \num{2.329}   & \num{0.000}   & \num{20.049}   \\
    \% change in Commission Revenue & \num{50.167} & \num{147.728} & \num{-97.497} & \num{1548.789} \\
    \bottomrule
    \end{tblr}
    \begin{flushleft}
        \small Note: This table shows the commission revenue that Airbnb earns as a whole platform for each market. Each value is expressed in units of $1/10,000$ yen. I computed commission revenue as the sum of each host's revenue multiplied by 0.03. The percentage change in commission revenue represents the extent to which commission revenue has changed from the Baseline scenario to the Self-Preferencing scenario.
    \end{flushleft}
\end{table} 

\setstretch{1.2}
\begin{table}[H]
    \centering
    \caption{Welfare Effects of Self-Preferencing}
    \label{Welfare_effects}
    \begin{tblr}[         
    ]                     
    {                     
    colspec={Q[]Q[]Q[]Q[]Q[]Q[]Q[]Q[]Q[]},
    cell{1}{2}={c=4,}{halign=c,},
    cell{1}{6}={c=4,}{halign=c,},
    cell{2}{2}={c=2,}{halign=c,},
    cell{2}{4}={c=2,}{halign=c,},
    cell{2}{6}={c=2,}{halign=c,},
    cell{2}{8}={c=2,}{halign=c,},
    column{1}={halign=l,},
    column{2}={halign=r,},
    column{3}={halign=r,},
    column{4}={halign=r,},
    column{5}={halign=r,},
    column{6}={halign=r,},
    column{7}={halign=r,},
    column{8}={halign=r,},
    column{9}={halign=r,},
    row{1}={halign=c,},
    row{2}={halign=c,},
    }                     
    \toprule
    & With Hotels & & & & Only Airbnb \\ \cmidrule[lr]{2-5}\cmidrule[lr]{6-9}
    & (i) RCNL &  & (ii) NL &  & (iii) RCNL &  & (iv) NL &  \\ \cmidrule[lr]{2-3}\cmidrule[lr]{4-5}\cmidrule[lr]{6-7}\cmidrule[lr]{8-9}
    & Mean & SD & Mean & SD & Mean & SD & Mean & SD \\ \midrule 
    CS (baseline)          & \num{20.704} & \num{5.567}  & \num{7.902}  & \num{2.139} & \num{13.360} & \num{3.232}  & \num{8.912}  & \num{4.802}  \\
    CS (self-preferencing) & \num{19.583} & \num{5.444}  & \num{7.788}  & \num{2.069} & \num{11.704} & \num{3.384}  & \num{8.231}  & \num{4.663}  \\
    \% change in CS       & \num{-5.077} & \num{7.856}  & \num{-0.985} & \num{4.907} & \num{-11.933} & \num{13.869}  & \num{-7.439} & \num{9.393}  \\
    \midrule
    PS (baseline)          & \num{0.756}  & \num{1.287}  & \num{0.281}  & \num{0.460} & \num{0.626}   & \num{0.964}  & \num{0.479}  & \num{0.573}  \\
    PS (self-preferencing) & \num{0.789}  & \num{1.294}  & \num{0.284}  & \num{0.461} & \num{0.706}   & \num{1.035}  & \num{0.527}  & \num{0.602}  \\
    \% change in PS       & \num{12.383} & \num{47.517} & \num{0.828}  & \num{3.494} & \num{30.962}  & \num{74.618} & \num{15.364} & \num{34.940} \\
    \midrule
    SW (baseline)          & \num{21.493} & \num{5.727}  & \num{8.183}  & \num{2.192} & \num{14.066}  & \num{3.357}  & \num{9.439}  & \num{4.841}  \\
    SW (self-preferencing) & \num{20.339} & \num{5.607}  & \num{8.072}  & \num{2.124} & \num{12.330}  & \num{3.495}  & \num{8.711}  & \num{4.703}  \\
    \% change in SW       & \num{-5.080} & \num{7.726}  & \num{-0.946} & \num{4.704} & \num{-11.936} & \num{13.522} & \num{-7.531} & \num{9.214}  \\
    \bottomrule
    \end{tblr}
    \begin{flushleft}
        \small Note: Each value is expressed in units of $1/10,000$ yen. CS, PS, SW, RCNL, and NL represent consumer surplus, producer surplus, social welfare, random coefficient nested logit, and nested logit, respectively. The percentage change in welfare represents the extent to which welfare has changed from the Baseline scenario to the Self-Preferencing scenario.
    \end{flushleft}
\end{table}
\doublespacing

\end{document}